\documentclass[11pt]{article} 

\PassOptionsToPackage{bbgreekl}{mathbbol}

\usepackage[showcomments]{ajtex}

\def\vol{\mathrm{vol}}
\def\half{\frac12}

\setlist[itemize]{itemsep=0pt,topsep=5pt}
\setlist[enumerate]{itemsep=0pt,topsep=5pt}

\usepackage{alphabeta}

\title{Higher-group global symmetry and the bosonic M5 brane}

\author{Jay Armas}\email{j.armas@uva.nl}
\author{Giorgos Batzios}\email{g.batzios@uva.nl}
\author{Akash Jain}\email{ajain@uva.nl}

\affiliation{Institute for Theoretical Physics, University of Amsterdam, 1090 GL Amsterdam, The Netherlands}
\affiliation{Dutch Institute for Emergent Phenomena, 1090 GL Amsterdam, The Netherlands}
\affiliation{Institute for Advanced Study, University of Amsterdam, Oude Turfmarkt 147, 1012 GC Amsterdam, The Netherlands}

\abstract{Higher-group symmetries are combinations of higher-form symmetries which appear in various field theories. In this paper, we explain how higher-group symmetries arise in 10d and 11d supergravities when the latter are coupled to brane sources. Motivated by this observation, we study field theories at zero and finite temperature invariant under a class of continuous Abelian higher-group symmetries.
We restrict the analysis to the low-energy regime where the dynamical field content exclusively consists of Goldstone fields arising from the spontaneous breaking of higher-group and spacetime symmetries.
Invariant quantities are constructed and the phases of matter are classified according to the pattern of spontaneous symmetry breaking. With respect to supergravity, we highlight how such Goldstone effective theories provide a symmetry-based interpretation for the theories living on D/M-branes. As an explicit example we construct a 6-group invariant action for the bosonic M5 brane, consistent with the self-duality of the 3-form field strength on the brane. While the self-duality condition in the bosonic case needs to be imposed externally as a constraint at zero temperature, we find an equilibrium effective action for the bosonic M5 brane at finite temperature that inherently implements self-duality.}

\begin{document}

\maketitle

\section{Introduction}
\label{introduction}

Symmetries provide a valuable organizing principle for developing effective field theories. A particularly successful example of this line of thinking is the Landau paradigm, which classifies phases of matter according to which global symmetries remain intact and which are spontaneously broken in the ground state. The advent of higher-form, or more generally categorical symmetries~\cite{Baez:2010ya, Gaiotto:2014kfa}, has dramatically broadened the scope of this paradigm, since it can incorporate theories with spatially extended conserved observables. While point particles are charged under an ordinary 0-form symmetry, the conserved objects charged under a $q$-form symmetry are $q$-dimensional branes/defects with $(q+1)$-dimensional worldvolume. The simplest example of this comes from electromagnetism, where the absence of magnetic monopoles implies that magnetic field lines cannot terminate and can be interpreted as 1-dimensional conserved strings associated with 1-form symmetry~\cite{Gaiotto:2014kfa, Grozdanov:2016tdf, Grozdanov:2017kyl, Hofman:2017vwr, Armas:2018atq, Armas:2018zbe}.
Just like 0-form symmetries, a $q$-form symmetry may also be spontaneously broken \cite{Lake:2018dqm, Hofman:2018lfz}. In fact, free electromagnetism without charged matter is understood as the spontaneously broken phase of the aforementioned 1-form symmetry, with the photon identified as the associated Goldstone boson. More examples include various topological phases of matter, see for instance~\cite{McGreevy:2022oyu} and references therein, including phases with emergent~\cite{Cherman:2023xok} and/or approximate higher-form symmetry~\cite{Armas:2023tyx}. In this paper we seek for an extension of the Landau paradigm to include non-trivial combinations of higher-form symmetries, namely higher-group symmetries~\cite{Cordova:2018cvg, Benini:2018reh, Iqbal:2020lrt, Brauner:2020rtz}. As a byproduct, this will allow us to propose a prescription for characterizing D/M-brane actions in 10d and 11d supergravities.

A continuous $q$-form symmetry may be represented by a continuous $(p+1)$-form current $J_{p+1}$ and, to probe the symmetry, consider its topological coupling 
with the background field $A_{q+1}$. The existence of symmetry is materialized in the conservation of the current, $\df{\star J_{q+1}}=0$, which is implemented at the level of the effective action via requiring its invariance under the background gauge transformation $\delta A_{q+1}= \df\Lambda_q$. 
If another higher-form symmetry is present, represented by the current $\mathcal{J}_{2q+2}$ and background field $B_{2q+2}$, the above conservation law can be modified to include a source term involving $\mathcal{J}_{2q+2}$.
This entanglement of the two higher-form symmetries can be implemented through the background transformation $\delta B_{2q+2} = \df\Lambda_{2q+1} - \kappa\,\df\Lambda_q \wedge A_{q+1}$ characterized by the higher-group structure constant $\kappa$. In general, there can be more than two higher-form symmetries participating in the higher-group structure, and we will briefly encounter such examples towards the end of this paper. Specializing to the case for which only two Abelian higher-form symmetries are present, the most general higher-group (modified) conservation laws featuring a scalar structure constant $\kappa$ take the form
\begin{align}\label{conservationlaw-intro}
    \df{\star J_{q+1}}
    &= 2\kappa\, G_{q+2}\wedge {\star \mathcal{J}_{2q+2}}~~, \nn\\
    \df{\star \mathcal{J}_{2q+2}} &= 0~~,
\end{align}
where $G_{q+2}=\df A_{q+1}$. This is a $(2q+2)$-group $U(1)^{(q)}\times_\kappa U(1)^{(2q+1)}$. The factor of 2 above results by appropriately choosing the background coupling of $\mathcal{J}_{2q+2}$ to ensure its gauge-invariance, as we explain in \cref{sec:higher-group-review} where we review higher-groups in more detail. It has been shown that theories enjoying higher-group symmetries arise in many contexts in quantum field theory \cite{Cordova:2018cvg,Benini:2018reh}. The prototypical examples are 2-group symmetries, $q=0$, that arise in a field theory with $U(1)^{(0)}\times U(1)^{(0)}$ direct product 0-form symmetry exhibiting a mixed 't Hooft anomaly, when one of the constituent 0-form symmetries is gauged~\cite{Cordova:2018cvg}. A useful analogy can be made with the Green-Schwartz mechanism~\cite{Green:1984sg} for anomaly cancellation that features similar transformation rules for dynamical gauge fields, though one should bear in mind that here $A_{q+1}$ and $B_{2q+2}$ are non-dynamical background fields.

This mixing of higher-form symmetries appears also in supergravities, in the low-energy effective description of branes. In particular, when one couples a supergravity theory to higher-form currents that can be carried by a probe brane, one obtains via Bianchi identities a set of conservation laws, whose complexity varies depending on the supergravity theory under consideration \cite{Armas:2016mes}.
As an example, for the case of the bosonic field content of 11d supergravity coupled to M2 and M5 brane sources with associated higher-form currents $J_{3}$ and $\mathcal{J}_{6}$, these equations are
\begin{align}\label{6-grouplaw1}
    \df{\star J_{3}}
    &=- G_{4} \wedge {\star \mathcal{J}_{6}} ~~, \nn\\
    \df{\star \mathcal{J}_{6}}
    &=0~~,
\end{align}
where $G_{4}=\df A_{3}$ is the 4-form invariant field strength of M-theory, which can be viewed as a fixed background field in the probe limit \cite{Armas:2016mes}. The source term in \cref{6-grouplaw1} has its origins in the Chern-Simons coupling $\int A_{3}\wedge G_{4} \wedge G_{4}$ contained in the 11d supergravity action. We note here that \cref{6-grouplaw1} can precisely be identified as a 6-group structure $U(1)^{(2)}\times_\kappa U(1)^{(5)}$ with the structure constant $\kappa = -1/2$. The M2 and M5 branes are charged objects under $U(1)^{(2)}$ and $U(1)^{(5)}$ parts of the 6-group symmetry, respectively. The 6-group structure indicates that M2 branes are sourced by M5 branes when coupled to a supergravity background with nonzero $G_4$ field strength. This physical application will be our primary driver in this work, but similar considerations also apply to D-branes in type IIA/B 10d supergravity theories, which we discuss in \cref{sec:outlook}. 
An important consequence is that the Goldstone modes associated with spontaneously broken higher-group symmetries must be given a supergravity interpretation. One might guess, correctly, that these Goldstone modes are none other than the usual vector and tensor modes appearing in brane (super)-multiplets.

In general, the spontaneous breaking of higher-group symmetries is characterized by the spontaneous breaking of the constituent higher-form symmetries to viable symmetry subgroups that may be respected by the ground state. At zero temperature, for higher-group symmetries involving two higher-form symmetries, i.e. $U(1)^{(q)}\times_\kappa U(1)^{(2q+1)}$, there is only one interesting phase where both the constituent higher-form symmetries are spontaneously broken, which we discuss in detail in \cref{sec:zero-T}. In principle, there can also be phases where the symmetry is spontaneously broken to the subgroup $U(1)^{(2q+1)}$, and another where it is unbroken. However, these are not very interesting from the low-energy point of view, because the latter has no low-energy degrees of freedom and the low-energy description of the former is identical to that of a $U(1)^{(q)}$ superfluid. Note that there is no phase where the residual symmetry of the ground state is $U(1)^{(q)}$ because it is not a proper subgroup of the $U(1)^{(q)}\times_\kappa U(1)^{(2q+1)}$ higher-group.\footnote{The reader may find an associated discussion for 2-groups in \cite{Cordova:2018cvg}.}

The phase space is comparatively richer at finite temperature, because the thermal frame of reference breaks the Lorentz symmetry and enlarges the set of potential subgroups that the ground state can be invariant under. This is a consequence of the qualitatively distinct feature of higher-form symmetries that they can partially spontaneously break in the time direction, while remaining unbroken in the spatial directions, called temporal spontaneous symmetry breaking~\cite{Armas:2018atq, Armas:2018zbe, Armas:2023tyx}. There are generically three phases of systems with $U(1)^{(q)}\times_\kappa U(1)^{(2q+1)}$ higher-group symmetry that have interesting low-energy dynamics: a T-T phase where both the higher-form symmetries are temporally spontaneously broken, a C-C phase where both are completely spontaneously broken, and a mixed C-T phase where the $U(1)^{(q)}$ symmetry is completely spontaneously broken but the $U(1)^{(2q+1)}$ symmetry is only temporally spontaneously broken. The converse symmetry breaking pattern, in which $U(1)^{(2q+1)}$ is completely spontaneously broken and $U(1)^{(q)}$ is only temporally spontaneously broken, is not allowed by the higher-group structure. We discuss these finite temperature phases in detail in \cref{sec:finiteT}.

The higher-form analogue of the Mermin-Wagner theorem forbids the spontaneous breaking of a $(2q+1)$-form symmetry in $d=2q+2,2q+3$ spacetime dimensions~\cite{Lake:2018dqm}. In critical dimensions $d=2q+3$, the classical effective field theory description is still qualitatively the same, except that the aspirant-Goldstones are massless fields that mediate quasi-long-range order with infinite correlation length instead of true long-range order characteristic of spontaneous symmetry breaking. In ``sub-critical'' dimensions $d=2q+2$, however, there is no such order at all and the aspirant-Goldstones are mere Stueckelberg fields with no dynamics. 
Consequently, there are only two finite-temperature phases in sub-critical dimensions: a T-U phase and a C-U phase, based on whether the $U(1)^{(q)}$ symmetry is temporally or completely spontaneously broken.
The $(2q+2)$-group symmetry structure is also quite trivial in this case because the current $\cJ_{2q+2}$, being a conserved top-form, is just proportional to the volume form $\star 1$. An interesting physical scenario that circumvents this triviality is when the theory under consideration is living on a dynamical defect itself propagating in an ambient higher-dimensional spacetime. In this case, the higher-group structure leaves imprints in the transverse dynamics and elastic fluctuations of the defect.

Supergravity provides a characteristic example of a higher-dimensional theory furnishing a higher-group symmetry on embedded geometries, as we further explain in the core of this paper. Returning to the M5 brane example for clarity, the theory on the worldvolume of this brane is an interacting theory of the $\cN = (2,0)$ massless super-Poincare multiplet. With respect to the bosonic field content, the existence of the brane itself spontaneously breaks translations in the 5 transverse directions giving rise to 5 scalar Goldstones. The remaining three bosonic degrees of freedom, required in order to have a match with those in the fermionic sector, come in form of a 2-form on the worldvolume with self-dual field strength. From a symmetry perspective, this 2-form is identified as the Goldstone mode associated with the spontaneously broken $U(1)^{(2)}$ part of the 6-group global symmetry \eqref{6-grouplaw1}. 
Note that the $U(1)^{(5)}$ part of the 6-group symmetry cannot be spontaneously broken on the 6-dimensional M5 worldvolume  because of the Mermin-Wagner theorem, and thus cannot give rise to Goldstone modes.
Interestingly, in this interpretation, all bosonic fields on the M5 worldvolume can be understood as Goldstone modes of spontaneously broken global symmetries. It is our expectation that the worldvolume fermions  can be treated on an equal footing using a combination of bosonic and fermionic higher-group symmetries.

At zero temperature, we proceed to construct a 6-group invariant effective action for the bosonic M5 brane, taking into account the self-duality of the 3-form field strength on its worldvolume. This is achieved by externally imposing the self-duality relation derived in \cite{Cederwall:1997gg} as a requirement of $\kappa$-invariance. We would like to emphasize that we do not write down a self-dual action, but rather an action consistent with self-duality, in the spirit of~\cite{Witten:1996hc,Cederwall:1997gg}. Unlike \cite{Cederwall:1997gg}, however, our effective action does not require a dynamical tension. Somewhat surprisingly, at finite temperature, it is possible to construct an equilibrium effective action that inherently implements the self-duality condition. In particular, we find that the bosonic M5 brane at finite temperature is described by a 6-group invariant theory in the T-U phase, where the constituent 2-form symmetry is temporally spontaneously broken but the 5-form symmetry is unbroken.

The plan for the rest of this paper is as follows. We begin our discussion with a lightening review of higher-group symmetries in \cref{sec:higher-group-review}. We then focus on the $(2q+2)$-group $U(1)^{(q)}\times_\kappa U(1)^{(2q+1)}$ and formulate a $(2q+2)$-group-invariant zero-temperature effective action in a phase where the symmetry is spontaneously broken in \cref{sec:zero-temp-action}. Along the way, we clarify subtleties with regards to the Mermin-Wagner theorem and when the theory is embedded in higher-dimensions. In \cref{sec:M5-brane} we study the specific application to M5 brane in 11d supergravity. \Cref{sec:finiteT} is devoted to higher-group symmetries at finite temperature. We set up the essentials of thermal partition functions in \cref{sec:thermal-pf}, followed by a detailed exposition of the various finite temperature phases in \cref{sec:phases}, and application to M5 brane in \cref{sec:M5-finite-T}. In \cref{sec:dual}, we briefly outline a dual formulation for the spontaneously broken phase, in terms of an anomalous higher-group symmetry involving four higher-form symmetries.
Finally in \cref{sec:outlook}, we close with potential extensions of our results and discuss research avenues for future exploration.

\section{(2q+2)-group symmetry at zero temperature}
\label{sec:zero-T}

In this section, we introduce the essentials of continuous Abelian higher-group symmetries that will be used throughout this paper. More details can be found in \cite{Cordova:2018cvg,Bhardwaj:2023kri}.
We primarily specialize to a $(2q+2)$-group, comprised of a $q$-form and a $(2q+1)$-form symmetry, and construct the effective action for a theory respecting this symmetry structure. We show that the worldvolume action of the bosonic M5 brane is a special case of a 6-group invariant action.

\subsection{Higher-group essentials}
\label{sec:higher-group-review}

A higher-group is defined as a collection of $p_{\sI}$-form symmetries which mix in a non-trivial manner. This mixing is encoded in the background gauge transformations of the corresponding background fields $A_{p_\sI+1}^{\sI}$.\footnote{In this paper we discuss higher-groups through the corresponding collection of background fields, and do not treat the group structure itself.} In addition to the usual transformation 
\begin{equation}
    \delta A_{p_\sI+1}^\sI = \df\Lambda_{p_\sI}^\sI~~,
\end{equation}
associated with a higher-form symmetry, a higher-group is characterized by the existence of at least one background gauge field which transforms as
\begin{equation}\label{highertrans}
    \delta A_{p_\sI+1}^\sI
    = \df \Lambda_{p_\sI}^\sI
    + \sum_{J} \Lambda_{p_\sJ}^\sJ \wedge {\mathfrak F}_{p_\sI+1-p_\sJ}(A)
    + \text{non-linear}~~,
\end{equation}
where the $(p_\sI+1-p_\sJ)$-form ${\mathfrak F}_{p_\sI+1-p_\sJ}$ is a function of (the products of) the background fields $A^\sK_{p_\sK+1}$, and their derivatives, with $p_\sK<p_\sI$. The transformation \eqref{highertrans} implies that the background field $A_{p_\sI+1}^\sI$ transforms under the transformation of the other lower-rank fields in the collection, in a way that couples the latter with a lower-rank gauge parameter. In general there can also be non-linear terms in the gauge parameters inside the transformation \eqref{highertrans}. Consequently, the associated field strength $F_{p_\sI+2}^\sI$ is no longer a closed form, but instead it is given by
\begin{equation}\label{fieldstrength}
    \mathcal{G}_{p_\sI+2}^\sI
    = \df A_{p_\sI+1}^\sI
    + {\mathfrak R}_{p_\sI+2}^\sI(A)~~,
\end{equation}
where ${\mathfrak R}_{p_\sI+2}$ is another function of the background fields $A_{p_\sK+1}^\sK$. Invariant field strengths given by equations of the form \eqref{fieldstrength} are the signatures of a higher-group symmetry. It is common in the literature to name a higher-group after the rank of the highest-rank background gauge field that transforms non-trivially under the higher-group gauge transformations. For instance, an interesting class of higher-groups is the Abelian $(n+1)$-group with a single background field, which we denote with $B_{n+1}$, transforming non-trivially, namely
\begin{equation}\label{n-group}
   \mathcal{H}^{(n+1)}= \left(\prod_{j} U(1)^{(p_{j})}\right) \times_{\kappa} U(1)^{(n)}~~,
\end{equation}
where the higher-group product $\times_{\kappa}$ is characterized by a single kind of structure constants collectively denoted by $\kappa$ which in general form a higher-dimensional symmetric matrix. Inside the product in \cref{n-group}, we have an ordinary direct-product structure of independent symmetries. As an explicit example of the structure in \cref{n-group}, consider the following 2-group gauge transformations
\begin{align}\label{B2many}
    \delta B_{2}
    &= \df\Lambda_{1}
    + \frac{1}{2\pi}\sum_{I,J}\kappa_{\sI\sJ}\,\Lambda_{0}^\sI\,\df A_{1}^\sJ\quad, \nn\\
    \delta A_{1}^\sI
    &= \df \Lambda_{0}^\sI~~,
\end{align}
with structure constants forming a symmetric matrix $\kappa_{\sI\sJ}$. This structure arises by gauging a flavor symmetry $U(1)^{(0)}_{C}$ in a parent theory with mixed $U(1)^{(0)}_\sI\times U(1)^{(0)}_\sJ\times U(1)^{(0)}_{C}$ t' Hooft anomaly~\cite{Cordova:2018cvg}. In this paper, we are mainly interested in a particular $(n+1)$-group which mixes two higher-form symmetries, but in \cref{sec:outlook} we will discuss a higher-group embedded in type IIB supergravity with multiple higher-form symmetries intertwining, whose structure is in fact more complicated than the one in \cref{n-group}.

In general, a continuous $p_\sI$-form symmetry can be represented by a $(p_\sI+1)$-form current $j^\sI_{p_\sI+1}$. Thus a continuous higher-group symmetry comes with a collection of currents $j_{p_\sI+1}^\sI$ and its local structure is completely determined by a set of (modified) conservation laws that these currents obey. At the same time, we can think of the field theory partition function $Z(A)$ as the functional integral $\int \cD\phi \exp(-iS(\phi,A))$ labeled by the background fields $A_{p_\sI+1}^\sI$, where $\phi$ denotes collectively the dynamical fields of the theory. Typically, the currents may be coupled to background fields in the effective action $S(\phi,A)$ via $\delta S\sim \int \delta A_{p_\sI+1}^\sI \wedge {\star j_{p_\sI+1}^\sI}$. However, for a higher-group, this coupling does not lead always to a gauge-invariant current. We can remedy this situation by allowing the coupling of each current with the background to include variations of other potentials in the collection, as we do below in an explicit example. 

Let us now turn our attention to the Abelian $(2q+2)$-group, 
$q \in N$, comprised of a $q$-form and a $(2q+1)$-form symmetry, with a single scalar\footnote{One might consider other more exotic cases where the structure constant itself is a fixed differential form but these are not relevant for our purposes here.} structure constant $\kappa$, i.e.
\begin{equation}\label{2l+2}
   H^{(2q+2)} 
    = U(1)^{(q)} \times_{\kappa} U(1)^{(2q+1)}~.
\end{equation}
The background transformations are 
\begin{align}
\label{2l+2group}
    \delta A_{q+1}
    &= \df\Lambda_q~~. \nn\\
    \delta B_{2q+2}
    &= \df\Lambda_{2q+1}
    - \kappa\, \df\Lambda_q\wedge A_{q+1}~~.
\end{align}
Here we have used a different, though equivalent, parametrization of the $B_{2q+2}$ background gauge transformation in comparison with \cref{B2many}, by moving the exterior derivative onto the gauge parameter in the $\kappa$-term. This amounts to a redefinition of the $(2q+1)$-form gauge parameter as $\Lambda_{2q+1}\rightarrow \Lambda_{2q+1} - \kappa\, \Lambda_{q}\wedge A_{q+1}$. The advantage of using \cref{2l+2group} will be made clear in \cref{sec:finiteT} where we introduce temperature. In addition to the $(q+2)$-form invariant field strength $G_{q+2}=\df A_{q+1}$, we can readily write down the $(2q+3)$-form invariant field strength
\begin{equation}
    \mathcal{G}_{2q+3}
    = \df B_{2q+2}+(-1)^{q+1}\kappa\, A_{q+1}\wedge G_{q+2}~~.
\end{equation}
Note that $\df\mathcal{G}_{2q+3}=(-1)^{q+1}\kappa\, G_{q+2}\wedge G_{q+2}$, so that the non-closedness of $\mathcal{G}_{2q+3}$ is present only when $q$ is even and, moreover, the non-closedness is proportional to the higher-group structure constant $\kappa$, as expected.
This further implies that the $(2q+2)$-group gets trivialized for odd values of $q$, and we can  simply redefine $B_{2q+2} \to B_{2q+2} - \kappa/2\,A_{q+1}\wedge A_{q+1}$ to decouple the two higher-form symmetries.
Given this, we consider $q$ to be even for the rest of this paper. 

Assigning to the $U(1)^{(q)}$ symmetry the current $J_{q+1}$ and to $U(1)^{(2q+1)}$ the current $\mathcal{J}_{2q+2}$, we may couple these symmetries to the classical sources $A_{q+1}$ and $B_{2q+2}$ respectively, through
\begin{equation}\label{firstcoupling}
    \delta S\sim 
    \int \delta A_{q+1} \wedge {\star J_{q+1}}
    + \Big( \delta B_{2q+2} + \kappa\, A_{q+1}\wedge \delta A_{q+1} \Big)
    \wedge {\star{\cal J}_{2q+2}}~~.
\end{equation}
Using this modified coupling for the higher-rank current $\mathcal{J}_{2q+2}$, we can define manifestly gauge-invariant currents via varying the effective action. It is visible from here as well that for odd values of $q$ we are led to 2 independent higher-form symmetries. Demanding invariance of the effective action under the $(2q+2)$-group gauge transformations \eqref{2l+2group} then produces the (modified) conservation laws
\begin{align}\label{conservationlaw}
    \df{\star J_{q+1}}
    &= 2\kappa\, G_{q+2}\wedge {\star \mathcal{J}_{2q+2}}~~, \nn\\
    \df{\star \mathcal{J}_{2q+2}} &= 0~~.
\end{align}
The higher-rank current $\mathcal{J}_{2q+2}$ still obeys a conventional conservation law while the conservation law of the lower-rank current $J_{q+1}$ is modified by a coupling of the current $\mathcal{J}_{2q+2}$ with the (closed) field strength $F_{q+2}$. Once we turn off the background fields, we recover two decoupled higher-form symmetries. Note that the factor of 2 on the right-hand side of $J_{q+1}$ conservation law in \cref{conservationlaw} is due to the gauge-invariant coupling in \cref{firstcoupling}, as was also observed in \cite{Iqbal:2020lrt}.

\subsection{(2q+2)-group invariant effective action}
\label{HGaction}
\label{sec:zero-temp-action}

Higher-form symmetries and their fusion into higher-groups arise in a plethora of gauge theories and (topological) field theories. They have been used to classify novel phases of matter and study phase transitions, extract constrains on symmetry-breaking energy scales, and shed new light on a broad spectrum of field theories; see e.g.~\cite{Gaiotto:2014kfa, McGreevy:2022oyu, Cordova:2022ruw, Cordova:2018cvg, Cordova:2022rer} and references therein. The simplest theories realizing higher-form or more generally higher-group symmetries are those for which one or more of the constituent higher-form symmetries are spontaneously broken. A reasonable first step is to build an effective field theory of dynamical higher-form Goldstone fields in the presence of higher-group invariance. In this section, we consider this effective theory for the $(2q+2)$-group $U(1)^{(q)} \times_{\kappa} U(1)^{(2q+1)}$, at zero temperature.

\subsubsection{Spontaneous symmetry breaking}
\label{sec:ssb}

A continuous $p$-form symmetry that is spontaneously broken in the ground state implies the existence of a massless Goldstone $p$-form field $\phi_{p}$ dominating the IR dynamics. The Goldstone fields typically transform with a phase, but for a higher-group this transformation rule might also need to include additional terms proportional to the structure constant $\kappa$.
In our case, we can spontaneously break the $U(1)^{(q)}$ part of the $(2q+2)$-group, giving rise to a $q$-form Goldstone field $\phi_{q}$ which transforms under \cref{2l+2group} as 
\begin{equation}\label{GolstoneTrans2}
    \delta \phi_{q}= -c_{\phi}\Lambda_{q}~.
\end{equation}
The parameter $c_{\phi}$ here counts the $q$-form charge of $\phi_q$, which we may always normalize to $1$ but we keep it unnormalized for now for clarity.
We can then write the following $(q+1)$-form that involves the exterior derivative of $\phi_{q}$, namely
\begin{equation}\label{invariant1}
    F_{q+1}=\df\phi_{q}+c_{\phi}A_{q+1}~~,
\end{equation}
and is invariant under \cref{2l+2group,GolstoneTrans2}.
Since $\phi_{q}$ is a dynamical field and all its dependence can only arise via the invariant $F_{q+1}$, the theory has an enhanced invariance under $\delta \phi_{q}=\df\lambda_{q-1}$, which is a true gauge symmetry. In that sense, $F_{q+1}$ is the invariant field strength associated to $\phi_{q}$. This reflects the fact that a $U(1)^{(q-1)}_{\text{local}}$ gauge theory is dual to a $U(1)^{(q)}$ superfluid. A well-explored example of this is $U(1)^{(0)}_{\text{local}}$ gauge theory, i.e. Maxwell's electromagnetism, which is dual to a $U(1)^{(1)}$ superfluid~\cite{Gaiotto:2014kfa, Armas:2018zbe}. The field strength 
$F_{q+1}$ satisfies the Bianchi identity
\begin{equation}\label{Bianchi3}
    \df F_{q+1}=c_{\phi}G_{q+2}~~.
\end{equation}

We now consider a phase where the $U(1)^{(2q+1)}$ part of the $(2q+2)$-group is spontaneously broken. The low-energy description then admits a $(2q+1)$-form Goldstone field $\Phi_{2q+1}$, with associated Goldstone charge $c_{\Phi}$. If one tries to construct a field strength associated to $\Phi_{2q+1}$ that is invariant under \cref{2l+2group}, one discovers the necessity of a $q$-form field that transforms by a shift of $\Lambda_{q}$. This is indeed the transformation law \eqref{GolstoneTrans2} of the $q$-form Goldstone $\phi_q$. To wit, we postulate the following transformation law for the $(2q+1)$-form Goldstone 
\begin{equation}\label{Golstonetrans5}
    \delta \Phi_{2q+1}=-c_{\Phi}\Lambda_{2q+1}
    - \kappa \frac{c_{\Phi}}{c_{\phi}} \df\phi_{q}\wedge \Lambda_{q}~~.
\end{equation}
As before, we may now write down the invariant field strength 
\begin{equation}\label{invariant2}
    \mathcal{F}_{2q+2}
    = \df\Phi_{2q+1}
    + c_{\Phi}B_{2q+2}
    + \kappa \frac{c_{\Phi}}{c_{\phi}} A_{q+1}\wedge \df\phi_{q}~~,
\end{equation}
which enjoys an additional invariance under a $2q$-form gauge symmetry $\delta \Phi_{2q+1}=\df \lambda_{2q}$. The Bianchi identity associated with $\mathcal{F}_{2q+2}$ is given as
\begin{equation}\label{Bianchi6}
    \df\mathcal{F}_{2q+2}
    = c_{\Phi}\mathcal{G}_{2q+3}
    + \kappa \frac{c_{\Phi}}{c_{\phi}}  G_{q+2}\wedge F_{q+1}~~.
\end{equation}
This construction implies that, generically, when the higher-rank symmetry of a $(2q+2)$-group is spontaneously broken, then the lower-rank symmetry must also be spontaneously broken. The same conclusion for a 2-group $U(1)^{(0)}\times_\kappa U(1)^{(1)}$ was obtained in~\cite{Cordova:2018cvg}, where it was argued that the $U(1)^{(0)}$ part of the 2-group is not a proper subgroup and thus cannot be a residual symmetry of the ground state after $U(1)^{(1)}$ is spontaneously broken.

Manifest invariance under \cref{2l+2group,GolstoneTrans2,Golstonetrans5} requires the effective action for Goldstone fields to take the general form  
\begin{equation}
\label{action}
    S=\int_{M} \star\cL(F_{q+1},\mathcal{F}_{2q+2})~~.
\end{equation}
Here $M$ is a $d$-dimensional (curved) Lorentzian manifold. Depending on the particular theory under consideration, the Lagrangian density $\cL$ can be constrained to take a particular form, for instance by imposing more symmetries.
As an example, we can write down the quadratic action
\begin{equation}
    S = -\int_{M} 
    g_{(q)} F_{q+1}\wedge \star F_{q+1}
    + g_{(2q+1)} \mathcal{F}_{2q+2}\wedge \star \mathcal{F}_{2q+2}~~,
\end{equation}
with arbitrary coupling constants $g_{(q)}$ and $g_{(2q+1)}$.
Strictly speaking, these effective actions are only valid in $d>2q+3$ spacetime dimensions because the higher-form analogue of the Mermin-Wagner theorem forbids the $U(1)^{(2q+1)}$ part of the symmetry to be spontaneously broken in $d=2q+3,2q+2$ dimensions, while there exists no higher-group of rank $m$ in spacetime dimensions $d<m$. Formally though, we can still make sense of this description, but with the Stueckelberg field $\Phi_{2q+1}$ not being interpreted as a Goldstone associated with any long-range order. In critical dimensions $d=2q+3$, $\Phi_{2q+1}$ mediates the so-called ``quasi-long-range order'', while in ``sub-critical dimensions'' $d=2q+2$, it is a mere auxiliary field with no dynamics and can be integrated out from the description. We discuss these scenarios in more detail in \cref{sub-critical}.

As a consistency check of the effective Goldstone action \eqref{action}, we must verify that the equations of motion of the two Goldstone fields are equivalent to the $(2q+2)$-group conservation laws. The equations of motion of $\phi_q$ and $\Phi_{2q+1}$ respectively read\footnote{We use the following definition for derivatives in the configuration space of p-forms: $\delta A(B) = \delta B \wedge \delta A/\delta B + \df(\ldots)$ for arbitrary differential forms $A$ and $B$.}
\begin{align}\label{phi-eom}
    \df\left(\frac{\delta{\star\cL}}{\delta F_{q+1}} \right)
    &= \kappa\frac{c_\Phi}{c_\phi}\, G_{q+2} \wedge 
    \frac{\delta{\star\cL}}{\delta \mathcal{F}_{2q+2}}~~, \nn\\
    \df\left(\frac{\delta{\star\cL}}{\delta \mathcal{F}_{2q+2}}\right)
    &=0~~.
\end{align}
Note that the equation of motion for $\phi_{q}$ receives a contribution due to the higher-group structure constant $\kappa$. 
The conserved currents can be computed by varying the action \eqref{action} with respect to the background gauge fields. Using the coupling structure in \cref{firstcoupling}, we can read out 
\begin{align}\label{eq:currents}
   {\star J_{q+1}}
   &= c_\phi\frac{\delta{\star\cL}}{\delta F_{q+1}}
   + \kappa\frac{c_\Phi}{c_\phi}\, F_{q+1} \wedge 
   \frac{\delta{\star\cL}}{\delta \mathcal{F}_{2q+2}}~~, \nn\\
   {\star\mathcal{J}_{2q+2}}
   &= c_\Phi\frac{\delta{\star\cL}}{\delta \mathcal{F}_{2q+2}}~~.
\end{align}
Given these expressions for the conserved currents, the equivalence between the Goldstone equations of motion \eqref{phi-eom} and the $(2q+2)$-group conservation laws \eqref{conservationlaw} follows immediately. Demanding the diffeomorphism-invariance of the action \eqref{action}, we are led to a modified conservation law for the energy momentum tensor $T^{\mu \nu}$, i.e.
\begin{equation}\label{eq:EM-conservation}
    \nabla_{\mu}T^{\mu \nu}
    = \frac{1}{(q+1)!}
    G_{q+2}^{\nu\mu_{1}\ldots\mu_{q+1}} 
    J_{q+1,\mu_{1}\ldots\mu_{q+1}}
    + \frac{1}{(2q+2)!}
    \cG_{2q+3}^{\nu\mu_{1}\ldots\mu_{2q+2}}
    \mathcal{J}_{2q+2,\mu_{1}\ldots\mu_{2q+2}}~~.
\end{equation}
Schematically, these equations take the same form as for two decoupled higher-form symmetries, except for the non-trivial higher-group structure implicit in the definition of $\cG_{2q+3}$.

\subsubsection{Mermin-Wagner theorem and (sub-)critical dimensions}
\label{sub-critical}

The ability of a charged operator to condense and spontaneously break a symmetry depends on the dimensionality $d$ of spacetime. For ordinary 0-form symmetries, the Mermin-Wagner theorem forbids spontaneous symmetry breaking in $d=1,2$. The statement can be generalized to higher $p$-form symmetries, which cannot be spontaneously broken in $d=p+1,p+2$ (while they do not exist in $d\leq p$)~\cite{Gaiotto:2014kfa, Lake:2018dqm}. In $d=p+2$, called critical dimensions, the theory can still admit a massless field $\phi_{p}$ but its fluctuations are so large that they destroy any long-range order, thereby restoring the symmetry in the ground state. This phenomenon is termed quasi-long-range order. 
While the quasi-long-range order cannot be distinguished based on the status of the $U(1)^{(p)}$ symmetry, it actually features a dual $U(1)^{(d-2-p)}$ symmetry associated with the Bianchi identity $\df{\star \tilde J_{d-1-p}} = 0$, where $\tilde J_{d-1-p} = \star\df\phi_p$ (in the absence of background fields), that is absent in the absence of long-range order. These phases are interpolated by a Berezinskii–Kosterlitz–Thouless (BKT) phase transition~\cite{1971JETP...32..493B, 1972JETP...34..610B, Kosterlitz:1973xp, 1979PhRvB..19.2457N, 1980PhRvB..22.2514Z}, mediated by topological defects in the configuration of $\phi_{p}$ that explicitly break the $U(1)^{(d-2-p)}$ symmetry~\cite{Armas:2023tyx}.

The Mermin-Wagner theorem still holds for each constituent $p$-form symmetry of a higher-group, and the $(2q+2)$-group structure imposes further constrains on the possible symmetry breaking patterns. Consequently, our discussion so far only strictly applies in $d>2q+3$. In the critical dimension $d=2q+3$ the massless field $\Phi_{2q+1}$ does not qualify as a Goldstone field because of the absence of any long-range order, but is nonetheless a dynamical field of the theory. Thus the action \eqref{action} can still be used for the critical dimension as well, but in this case the invariant field strength $\mathcal{F}_{2q+2}$ has no longer the interpretation of a Goldstone field strength.

In sub-critical dimensions $d=2q+2$, there cannot be a notion of a massless $\Phi_{2q+1}$ field at all. Note that the associated field strength $\cF_{2q+2}$ is a top-form, hence all the gauge-invariant (physical) information in $\Phi_{2q+1}$ can be encoded into a single unconstrained scalar field $X=\star\cF_{2q+2}$. The action $S$ of the theory generically admits a term $\cF_{2q+2}\wedge{\star\cF_{2q+2}} = -{\star 1} X^2$, which acts as a mass term for $X$ gapping it out from the low-energy spectrum. The resultant action takes a generic form
\begin{equation}
    \label{eq:action-subcritical}
    S = \int_M \star\hat\cL(F_{q+1})
    + Q_{(2q+1)}\int_M \cF_{2q+2}~~,
\end{equation}
for a constant $U(1)^{(2q+1)}$ charge $Q_{(2q+1)}$.
Note that the occurrence of $\Phi_{2q+1}$ within $\cF_{2q+2}$ is purely a boundary term that drops out from the theory provided that the manifold $M$ is unbounded. It has no dynamics and its only role here is to make the Lagrangian manifestly gauge-invariant. Note that the associated $U(1)^{(2q+1)}$ current is given by $\star\cJ_{2q+2} = Q_{(2q+1)}$, which trivially satisfies the conservation equation $\df{\star\cJ_{2q+2}} = \df Q_{(2q+1)} = 0$. The $U(1)^{(q)}$ current, on the other hand, is given as
\begin{equation}
    {\star J_{q+1}} 
    = c_\phi\,\frac{\delta{\star\hat\cL}}{\delta F_{q+1}}
    + Q_{(2q+1)}\frac{\kappa}{c_\phi} F_{q+1}~~,
\end{equation}
which satisfies the (non-)conservation equation derived from \cref{conservationlaw}, i.e.
\begin{align}
    \label{eq:anomalous-q}
    \df{\star J_{q+1}}
    &= 2\kappa Q_{(2q+1)} G_{q+2}~~.
\end{align}
Interestingly, the source term in this equation due to the $U(1)^{(q)}\times_\kappa U(1)^{(2q+1)}$ higher-group structure can effectively be seen as a $U(1)^{(q)}$-invariant theory with a 't Hooft anomaly $\kappa Q_{(2q+1)}$. This can be made more precise in the context of the anomaly inflow mechanism, by coupling the theory to a $(d+1)$-dimensional bulk manifold carrying a $q$-form Chern-Simons theory 
\begin{align}\label{eq:bulk-action}
    S_{\text{bulk}} 
    &= - Q_{(2q+1)} \int\mathcal{G}_{2q+3}~~ \nn\\
    &= \int_{\text{bulk}} \kappa Q_{(2q+1)}\, A_{q+1}\wedge \df A_{q+1}
    - Q_{(2q+1)} \df B_{2q+2} ~~.
\end{align}
Since this action only depends on the background fields, it does not change the dynamics of the original theory. The dependence on $B_{2q+2}$ identically cancels between $S$ and $S_{\text{bulk}}$, and thus the total theory described by $S + S_{\text{bulk}}$ is $U(1)^{(q)}$-invariant without invoking the higher-group structure, but with an anomaly inflow from $S_{\text{bulk}}$. The higher-group structure in sub-critical dimensions is non-trivial if the theory is defined on a $(2q+2)$-dimensional brane/defect propagating in a higher-dimensional spacetime. We will discuss this scenario in \cref{sec:embedding}.

As we argued above, the auxiliary Stueckelberg field $\Phi_{2q+1}$ can always be integrated out from the low-energy description. Nevertheless, it is convenient to keep it in the formulation of the effective theory and work with the action \eqref{action} even in sub-critical dimensions $d=2q+2$. Of course, this field has no low-energy dynamics but it allows us to write down Lagrangians that are manifestly gauge-invariant, as opposed to invariant up to boundary terms. It also keeps the discussion manifestly analogous to higher dimensions $d> 2q+2$. An incarnation of $\Phi_{2q+1}$, for $q=2$, also appeared in the supergravity literature in the context of an M5 brane action~\cite{Cederwall:1997gg}; we will discuss this in more detail in \cref{sec:M5-brane}.

\paragraph*{(Anti-)self-dual theories:} Another qualitatively distinct feature of $(2q+2)$-group-invariant theories in sub-critical dimensions $d=2q+2$ is that the $U(1)^{(q)}$ Goldstone $\phi_q$ can be constrained to be (anti-)self-dual. Linearly, this constraint is
\begin{equation}
    \star F_{q+1} = \pm F_{q+1} + \ldots~~.
    \label{eq:self-duality}
\end{equation}
Recalling that $\phi_q$ realizes a $U(1)^{(q-1)}_{\text{local}}$ gauge symmetry, this is the familiar (anti-)self-duality constraint of $(q-1)$-form gauge theories in $2q+2$  dimensions. Taking another Hodge-dual of this equation, one can verify that this constraint can only be satisfied for $F_{q+1}\neq 0$ provided that $q$ is even, as is the case for us. To understand the meaning of (anti-)self-duality constraint non-linearly, note that such a constraint will generically make the system of dynamical equations over-constrained unless the $q$-form conservation equations \eqref{eq:anomalous-q} were identical to the Bianchi identities \eqref{Bianchi3}. This allows us to straightaway read out the non-linear (anti-)self-duality relation as
\begin{equation}
    \star F_{q+1} = \frac{c_\phi}{2\kappa Q_{(2q+1)}} J_{q+1}~~.
    \label{eq:non-linear-self-duality}
\end{equation}
However, not every sub-critical $(2q+2)$-group theory described by the effective action \eqref{eq:action-subcritical} is compatible with this (anti-)self-duality constraint. 
Consider, for example, ${\star\hat\cL} = \half f F_{q+1}\wedge \star F_{q+1}$, where $f$ is a constant. It immediately follows that 
\begin{equation}
    J_{q+1} = c_\phi f F_{q+1}
    + Q_{(2q+1)}\frac{\kappa}{c_\phi}\, {\star F_{q+1}}
    \quad\overset{\text{self-duality}}{\implies}\quad 
    \star F_{q+1}
    = \frac{c_\phi^2 f}{\kappa Q_{(2q+1)}} F_{q+1}
    ~~,
\end{equation}
which can only be satisfied for $F_{q+1}\neq 0$ provided that $f = \pm \kappa Q_{(2q+1)}/c_\phi^2$. 
In other words, to define non-trivial (anti-)self-dual theories, we must also require that the Hodge-dual of \cref{eq:non-linear-self-duality} is identically satisfied upon reusing \cref{eq:non-linear-self-duality} to eliminate $\star F_{q+1}$. In detail, we must demand that
\begin{equation}
    F_{q+1} = 
    \frac{c_\phi}{2\kappa Q_{(2q+1)}}\,{\star J_{q+1}}
    \bigg|_{\star F_{q+1}\to \frac{c_\phi}{2\kappa Q_{(2q+1)}} J_{q+1}~\text{recursively}}~~,
    \label{eq:hodge-dual-constraint}
\end{equation}
holds as an identity. One can check that this applies to our simple example above. We will see a more non-trivial example of a consistent self-dual theory in \cref{sec:M5-brane} during our discussion of M5 branes in supergravity.

\subsubsection{Embedded geometries}
\label{sec:embedding}

In this subsection, we consider a higher-group-invariant field theory as being embedded in a higher-dimensional background theory, which furnishes the higher-group symmetry. To be concrete consider a brane, or more generally a defect, with embedding functions $X^{\mu}(\sigma)$ in the ambient background spacetime with metric $g_{\mu \nu}$, where $\sigma^{a}$ are taken to be coordinates on the worldvolume $M$. We impose diffeomorphism invariance among the $\sigma^{a}$ coordinates on $M$, leaving $D-d$ independent degrees of freedom in $X^{\mu}(\sigma)$ parametrising the elastic fluctuations of the brane in the transverse directions. One can define an induced metric $\gamma_{a b}=\partial_{a} X^{\mu} \partial_{b}X^{\nu}g_{\mu \nu}$ as the pullback of the background metric $g_{\mu \nu}$. More details on calculus of embedded geometries can be found in~\cite{Carter:2000wv, Emparan:2009at}.

Our attention shall be restricted to a $ U(1)^{(q)}\times_\kappa U(1)^{(2q+1)}$ global symmetry, acting across the whole spacetime. The corresponding background fields $\breve A_{q+1}$, $\breve B_{2q+2}$ and currents $\breve J_{q+1}$, $\breve\cJ_{2q+2}$ are defined on the fixed background spacetime $\breve M$, with pullbacks onto the dynamical worldsheet $M$ denoted by $A_{q+1}$, $B_{2q+2}$ and $J_{q+1}$, $\cJ_{2q+2}$ respectively. Here, we use the ``breve'' accent to denote quantities living on $\breve M$ and the regular versions for those living on $M$. In particular, note that the $A_{q+1}$, $B_{2q+2}$ are not purely background fields in this setting and contain the dynamical fields $X^{\mu}$ as well through pullback maps. We emphasize again that there is only a single higher-group symmetry that acts on all of $\breve M$. The higher-group structure on $M$ is inherited from $\breve M$ through pullback maps.

Then, the low-energy description of the defect is an interacting theory of the Goldstone fields associated with spontaneously broken translations as well as spontaneously broken higher-form symmetries. The former are identified with the $D-d$ scalar degrees of freedom contained within the transverse sector of $X^\mu(\sigma)$. In general, the $(2q+2)$-group symmetry may be spontaneously broken in all of spacetime $\breve M$, giving rise to the Goldstone fields $\phi_{q}$ and $\Phi_{2q+1}$ having legs on directions tangential as well as perpendicular to $M$. We further restrict the analysis to a simpler physical scenario where the source of spontaneous symmetry breaking is confined to the worldvolume of the defect, forcing the Goldstones to have only tangential indices. Their transformation rules are still given by \cref{GolstoneTrans2,Golstonetrans5}, but with background gauge fields and higher-form symmetry parameters pulled back onto $M$.

The invariant coupling of the defect theory with the background spacetime is provided by the invariant field strengths defined in \cref{invariant1,invariant2}, where the background fields are
understood as being pulled-back on $M$. As a result, the effective action for the defect is parameterized as in \cref{action}, but with the dynamical embedding fields $X^\mu$ appearing implicitly within the invariants $F_{q+1}$, $\cF_{2q+2}$ and in the volume-form $\vol_d$ used to define the integrals and Hodge-duality operation on $M$. The conserved currents $J_{q+1}$, $\cJ_{2q+2}$ on the defect can be obtained by varying the action with respect to pullback gauge fields $A_{q+1}$, $B_{2q+1}$, and take the same form as in \cref{eq:currents}. Since all dependence on the spacetime background fields $\breve A_{q+1}$, $\breve B_{2q+1}$ arise implicitly via $A_{q+1}$, $B_{2q+1}$ in the action, the associated conserved currents $\breve J_{q+1}$, $\breve\cJ_{2q+2}$ are simply given by pushforward of worldvolume currents onto $\breve M$, multiplied with delta functions localized on $M$. The spacetime currents $\breve J_{q+1}$, $\breve\cJ_{2q+2}$ satisfy the $(2q+2)$-group conservation equations analogous to \cref{conservationlaw}, with the $(q+2)$-form field strength $\breve G_{q+2}$. The components of these conservation equations transverse to $M$ are identically satisfied because the respective conserved currents themselves do not carry any nonzero transverse components. On the other hand, the projection of these equations onto $M$ lead to the conservation equations for $J_{q+1}$, $\cJ_{2q+2}$, taking the same form as \cref{conservationlaw}. However, note that the occurrence of $G_{q+2}$ in these equations contains the dynamical fields $X^\mu(\sigma)$ within its definition.

An intriguing case to consider is when the defect is $(2q+2)$-dimensional, i.e. it is sub-critical with respect to the $(2q+2)$-group structure. Naively, following our discussion in \cref{sub-critical}, one might expect that the $(2q+2)$-group structure in this case may also be effectively interpreted as a $U(1)^{(q)}$ invariant theory with a 't Hooft anomaly $\kappa Q_{2q+1}$. After all, a defining property of the higher-group is the existence of at least one non-closed background field strength, in our case $\mathcal{G}_{2q+3}$, and such a differential form identically vanishes in $d=2q+2$ dimensions. Nevertheless, $\mathcal{G}_{2q+3}$ appears explicitly in the total energy-momentum conservation equation for the defect. This equation is identical to \cref{eq:EM-conservation}, where all indices are understood as spacetime, rather than worldvolume. While the projection of this equation along the directions tangential to the worldvolume, dictating the intrinsic dynamics of the defect, has no sign of $\mathcal{G}_{2q+3}$, the projection transverse to the defect directions leads to
\begin{equation}
    T^{a b}K^{\ \ I}_{a b}
    = \frac{1}{(q+1)!}G_{q+2}^{Ia\ldots}\,J_{q+1,a\ldots}
    + \frac{1}{(2q+2)!}\mathcal{G}_{2q+3}^{Ia\ldots}\,\mathcal{J}_{2q+2,a\ldots}~~,
    \label{eq:elastic-eqn}
\end{equation}
which determines the extrinsic dynamics of the defect in the ambient spacetime.
In writing this equation, we have introduced the set of normal vectors $n^I_\mu$ transverse to $M$, satisfying the relations $n_\mu^I t_a^\mu = 0$ and $n_\mu^I n_\nu^J g^{\mu\nu} = \delta^{IJ}$, where $t^\mu_a = \dow_a X^\mu$ are the tangent vectors on $M$. Using these, we have further defined the the mixed components of the higher-form background field strengths $G_{q+2}^{Ia\ldots} = n^I_\lambda t^a_\mu\ldots \breve G_{q+2}^{\lambda\mu\ldots}$ and $\cG_{2q+3}^{Ia\ldots} = n^I_\lambda t^a_\mu \ldots \breve\cG_{3q+3}^{\lambda\mu\ldots}$, together with the second fundamental (or extrinsic curvature) tensor $K^{\ \ I}_{a b}=n^I_{\mu}\nabla_{a}t_b^{\mu}$. Therefore, we observe that the non-closed field strength affects the extrinsic dynamics of the defect.

This points towards the fact that, in this case, the $(2q+2)$-group structure is non-trivialized by the propagating nature of the defect in higher dimensions. Indeed, the source term in \cref{eq:anomalous-q} for an embedded defect cannot be interpreted as an anomaly because it contains the dynamical fields $X^\mu(\sigma)$ implicitly within the definition of the background field strengths. If we were to try and setup an anomaly inflow mechanism, the $(d+1)$-dimensional bulk action in \cref{eq:bulk-action} would also contain the dynamical fields $X^\mu(\sigma)$, thereby modifying the dynamics of the theory. We infer that the non-trivial higher-group structure is inherited from the higher-dimensional theory via the elastic fluctuations of the defect.

\subsection{M5 brane action}
\label{sec:M5-brane}

The framework developed in the previous subsections can naturally accommodate the classical low-energy dynamics of string theory solitons. D- and M-branes are half-BPS objects and as such carry a supersymmetric field theory on their worldvolume. The low-energy field content should then correspond to a massless super-Poincare multiplet.
Working in the supergravity approximation, the effective action of M-theory and type II string theories is effectively approximated by $S=S_{\text{sugra}}+S_{\text{brane}}$. Constraining ourselves to the dynamics of the brane embedded in an on-shell background, it is reasonable to expect that the supergravity origin of these massless modes is associated to spontaneously broken symmetries of the background. In other words, there is a Goldstone mechanism in action. It was understood a long time ago that the scalars contained in each brane multiplet must be identified with the Goldstone modes arising from the spontaneous breaking of translational invariance in directions transverse to the brane~\cite{Polchinski:1996na, Johnson:2000ch}. 
The 8 propagating fermionic degrees of freedom should then be understood as Goldstone modes of spontaneously broken supersymmetries. At the same time, the Goldstone nature of the vector and antisymmetric tensor modes has remained somewhat unclear. 

If we consider the background gauge fields of 11d supergravity and type IIA/B 10d supergravities that are coupled to M- and D-branes respectively, then from their transformation properties we infer for each of these cases a higher-group global symmetry whose charged objects are M- and D-branes respectively. Equivalently, this can be derived by inspection of the corresponding conservation laws. We thus propose that the vector and tensor modes on a brane can be understood as the Goldstone fields coming from the spontaneous breaking of higher-group symmetries. To make these ideas concrete below we apply them to the M5 brane of 11d supergravity. We discuss the higher-group perspective for other branes in \cref{sec:outlook}.

There are 5 transverse directions to the M5 brane, giving rise to 5 scalar degrees of freedom. Supersymmetry implies that we need 3 more propagating bosonic degrees of freedom in order to match with the 8 fermionic ones. Not surprisingly, there exists a unique tensor multiplet in $d=6$ dimensions involving 5 scalars, and in addition it contains a 2-form $V_{2}$ with self-dual field strength, which provides exactly the 3 required bosonic degrees of freedom.
Thus the theory on the M5 brane must be an interacting theory of the $\mathcal{N}=(2,0)$ multiplet. In this work we are mainly interested on an action principle for the bosonic field content, though we believe that the worldvolume fermions can be treated on an equal footing using a combination of bosonic and fermionic higher-form symmetries. Focusing on the bosonic field content, the coupling of the M-theory equations of motion to the brane sources leads to the conservation laws in \cref{6-grouplaw1}. These are immediately identified with the $(2q+2)$-group conservation laws \eqref{conservationlaw} where $q=2$ and $\kappa=-1/2$. We shall refer to this symmetry as the 6-group of 11d supergravity, given by $U(1)^{(2)}\times_\kappa U(1)^{(5)}$. The corresponding background fields $A_{3}$ and $B_{6}$ are the potentials for the M-theory spacetime invariant field strengths $G_{4}$ and ${\cal G}_{7}$, respectively.\footnote{The 11d supergravity solution further forces $\cG_7 = \star_{11}G_4$.} The 5-form topological symmetry $U(1)^{(5)}$, whose charged operators are M5 branes, says that the 5-brane charge is conserved, while the $U(1)^{(2)}$ symmetry is a topological symmetry with M2 branes acting as the charged operators. The latter is assigned a modified conservation law arising from the underlying 6-group structure, which, from the perspective of the M5 worldvolume, dictates the distribution of the M2 brane charge within (or diluted in) the M5 brane.

Irrespective of the glaring issue of self-duality, if one now tries to couple the bosonic M5 brane to the (pulled-back) M-theory background $A_{3}, B_{6}$, then one finds that the 2-form must transform as 
\begin{equation}
    \delta V_{2}=-\Lambda_{2}~~,
\end{equation}
when $\delta A_{3}=\df\Lambda_{2}$ and $\delta B_{6}=\df\Lambda_{5} + 1/2\,\df\Lambda_{2}\wedge A_{3}$. This is precisely the transformation law of a 2-form Goldstone in \cref{GolstoneTrans2}, implying that we must identify $V_{2}=\phi_{2}$ (normalized with unit charge $c_\phi=1$), i.e. the 2-form on the M5 worldvolume is the Goldstone field arising from the spontaneous breaking of the $U(1)^{(2)}$ part of the 6-group $U(1)^{(2)}\times_{\kappa} U(1)^{(5)}$. This already tells us that the action of the bosonic M5 brane in a fixed background should be understood as a 6-group invariant effective action for the propagating Goldstone mode $\phi_{2}$. 

To manifest the 5-form part of the 6-group symmetry, it will also be convenient to introduce the 5-form Stueckelberg field $\Phi_5$ on the worldvolume that transforms as $\delta \Phi_5= - \Lambda_5 + \half \df\phi_2 \wedge \Lambda_2$, with $c_\Phi = 1$. As we discussed in \cref{sub-critical}, this field has no dynamical degrees of freedom on a 6-dimensional worldvolume. The fact that $\Phi_{5}$ is redundant and can be removed completely from the description fits nicely with the field content of the $\mathcal{N}=(2,0)$ multiplet, where no such field exists.
The introduction of this non-dynamical field in the M5 brane action first appeared in \cite{Cederwall:1997gg}, motivated by the proposal of \cite{Cederwall:1997ab}. From our perspective, at low energies, each dynamical worldvolume field on the brane is associated to the spontaneous breaking of a higher-form global symmetry.

Given the results in the previous subsections, the invariant coupling of the brane to the background is provided through the field strengths $F_{3}$ and $\mathcal{F}_{6}$ defined in \cref{invariant1,invariant2} respectively. We emphasize that this 6-group symmetry is non-trivial, despite of the M5 worldvolume being 6-dimensional, because of its ability to move and bend inside the 11-dimensional background spacetime and thus should be used as an organizing principle for the theory on the bosonic M5 brane. 

The self-dual nature of the field strength $F_{3}$ on the worldvolume of the M5 brane
raises obstructions in the formulation of an action principle. The root of the problem is traced back to the simple fact that at the linearized level the kinetic term $F_{3}\wedge \star F_{3}$ vanishes. Nevertheless, one can still search for an action which produces the self-duality constraint on-shell. There are a few approaches in the literature to achieve this (see for instance \cite{Simon:2011rw} and references therein), of which we highlight~\cite{Perry:1996mk} that singles out a spatial direction thus losing manifest worldvolume covariance, as well as \cite{Pasti:1997gx} where the worldvolume covariance is restored at the expense of introducing a single auxiliary field, furnishing an extra gauge symmetry. Another route was proposed in \cite{Witten:1996hc}, according to which one writes an action that is consistent with self-duality, without directly yielding it. The self-duality constraint is then implemented at the level of the path integral. This approach was extended classically at the fully non-linear level in \cite{Cederwall:1997gg} using a dynamical brane tension. We view this last approach as the most suitable when the theory on the brane is formulated as an effective theory of spontaneously broken global higher-form symmetries.

As mentioned earlier, in order to retain manifest invariance under the 6-group transformations of the background gauge fields, the action must depend solely on the invariants $F_{3}$ and $\mathcal{F}_{6}$. The action for the bosonic Abelian M5 brane takes the form
\begin{equation}\label{M5action}
    S = -Q_{(5)}\int_{M} \star \sqrt{1+\mathcal{Y}(F_{3})}
    + Q_{(5)}\int_M \cF_{6}~~,
\end{equation}
where $Q_{(5)}$ is the M5 brane charge, proportional to the tension $T_{\text{M5}}$ of the brane, and $\cY(F_3)$ is a scalar function of $F_3$.
The dependence of this action on the transverse scalars is inherited through the induced metric $\gamma_{ab}$ on the worldvolume $M$ of the M5 brane. 
Here and for the rest of this section the Hodge dual operator $\star$ is taken with respect to the induced metric $\gamma_{ab}$.
It is important to stress that at this point $\phi_{2}$ is an unconstrained 2-form, the reasoning being that we embed the chiral (self-dual) theory into a non-chiral theory. At the classical level, the chiral sector is extracted by demanding the equations of motion of $\phi_{2}$, i.e.
\begin{equation}
     \df{\star\!\left(
     \frac{1}{\sqrt{1+\mathcal{Y}(F_{3})}}\frac{\delta \mathcal{Y}(F_3)}{\delta F_{3}}
     \right)}
     =G_{4}~,
\end{equation}
to be equivalent to the Bianchi identity $\df F_{3}=G_{4}$, thereby determining the self-duality condition. See our discussion around \cref{eq:self-duality}.
Using the equations of motion it is straightforward to see that the self-duality relation reads
\begin{equation}
    \star F_{3} = \frac{1}{\sqrt{1+\cY(F_3)}} \frac{\delta \cY(F_3)}{\delta F_3}~~,
    \label{eq:general-M5-selfduality}
\end{equation}
where the right-hand side is understood as a function of $F_{3}$. We point out that invariance under 6-group symmetry is not enough to further constrain $\cY(F_3)$, so as to give rise to a consistent self-dual theory. Supersymmetry is responsible for the the matching between bosonic and fermionic degrees of freedom, and one way to proceed is to derive the explicit form of the self-duality relation by requiring the supersymmetric extension of the action \eqref{M5action} to be $\kappa$-invariant. Indeed, it was shown in \cite{Cederwall:1997gg} that in order for the brane configuration to retain supersymmetry, a relation that expresses $\star F_{3}$ as a non-linear function of $F_{3}$ must hold. In what follows, we obtain an effective action for the bosonic M5 brane that will be proven to be consistent with the aforementioned self-duality relation.

Starting from the form of the action \eqref{M5action}, the specific scalar function $\mathcal{Y}(F_{3})$ in \eqref{M5action} compatible with self-duality is given by
\begin{equation}
     \mathcal{Y}(F_{3})
     = \frac{1}{12}F_{abc}F^{abc}
     + \frac{1}{288}({F_{abc}F^{abc})^2}-\frac{1}{96}\left(F_{abc} F^{hbc}F_{hde} F^{ade}\right)~.
\end{equation}
Using this into \cref{eq:general-M5-selfduality}, the self-duality condition takes the form\footnote{The square brackets denote antisymmetrization over indices, i.e. $X_{[a_1\ldots a_n]} = \frac{1}{n!} \lb X_{a_1\ldots a_n} + \text{signed permutations}\rb.$}
\begin{equation}\label{self-dualexplicit}
    \star F_{abc}=\frac{1}{\sqrt{1+\mathcal{Y}(F_{3})}} 
    \left(F_{abc}
    + \frac{1}{12} F_{deh}F^{deh}F_{abc}
    - \frac{1}{4} F_{de[a} F_{bc]h} F^{hde}\right)~~.
\end{equation}
This is exactly the self-duality condition obtained in \cite{Cederwall:1997gg} as a requirement of $\kappa$-invariance. This completes the construction of the action of the bosonic M5 brane as a Goldstone effective action invariant under the 6-group $U(1)^{(2)}\times_\kappa U(1)^{(6)}$, which is consistent with the self-duality property of the 2-form $\phi_{2}$.

Before closing this section, we should emphasize that \cref{M5action} is not a self-dual action, i.e. the self-duality relation does not arise as an equation of motion. At the level of bosonic degrees of freedom, self-duality is an extra input that we essentially impose by hand. We point out, however, that this relation should be extractable as a requirement of an enhanced global symmetry once one takes into account the worldvolume fermions. The action \eqref{M5action} is consistent with (non-linear) self-duality, since imposing the latter forces the 2-form equation of motion to be merely an identity, as explained in \cite{Witten:1996hc, Cederwall:1997gg}. Nonetheless, this is enough to predict the full non-linear interactions in the Lagrangian.

\section{(2q+2)-group symmetry at finite temperature}
\label{sec:finiteT}

The purpose of this section is to study field theories with $(2q+2)$-group symmetry at finite temperature, and classify equilibrium phases with different spontaneous symmetry breaking patterns depending on the continuous subgroups of $U(1)^{(q)}\times_{\kappa} U(1)^{(2q+1)}$ respected by the equilibrium state.
The configuration equations and constitutive relations are derived for all the allowed equilibrium phases. In particular, we identify the M5 brane phase at finite temperature based on 6-group symmetry and self-duality.

\subsection{Thermal partition function}
\label{sec:thermal-pf}

In this work we restrict the analysis to the description of phases with $(2q+2)$-group symmetry in thermal equilibrium. At finite temperature, the Lorentz symmetry of the theory is broken by the existence of the thermal observer $\beta^{\mu}=\delta^{\mu}_{t}/T_0$, with $T_0$ being the constant equilibrium temperature. The background fields are taken to be time-invariant with respect to the thermal observer, i.e.
\begin{equation}
    \lie_\beta A_{q+1} = \lie_\beta B_{2q+2} = \lie_\beta g_{\mu\nu} = 0~~,
\end{equation}
where $\lie_{\beta}$ denotes the Lie derivative with respect to $\beta^\mu$. In other words, $\beta^\mu$ is a Killing vector for the spacetime background.
We can express the $d$-dimensional (sub-)manifold on which the theory resides as $M=\mathcal{R}\times \Sigma$, where $\cR$ denotes the direction of time along $\beta^{\mu}$ and $\Sigma$ denotes a $(d-1)$-dimensional spatial Cauchy slice transverse to $\beta^{\mu}$. The fundamental object of interest at finite temperature is the thermal partition function $\cZ$, defined as a path integral over time-independent configurations of the dynamical fields collectively denoted as $\phi$, namely 
\begin{equation}\label{partitionf}
    \mathcal{Z}[A_{q+1},B_{2q+1},g_{\mu\nu}]
    = \int\cD\phi\,
    {\rm e}^{-S_{E}\lb\phi;A_{q+1},B_{2q+1},g_{\mu\nu}\rb}~~,
\end{equation}
with
\begin{equation}
    \lie_\beta \phi = 0~~,
\end{equation}
where $S_{E}$ is the Euclidean effective action, defined as an integral over 
$\Sigma$. Note that $\cZ$ is a functional of the time-independent configurations of the background fields $A_{q+1}$, $B_{2q+2}$, and $g_{\mu\nu}$.
The thermal expectation values and correlation functions of the higher-form currents and the energy-momentum tensor are obtained as usual by varying $\cZ$ with respect to these background fields. 

The precise composition of the dynamical fields $\phi$ depends on the equilibrium phase under consideration and will generically contain different space- and time-components of the Goldstone fields $\phi_q$ and $\Phi_{2q+1}$. From the point of view of hydrodynamics, we are essentially describing different phases of a $(2q+2)$-group superfluid in thermal equilibrium. These are connected to the fact that the class of continuous subgroups of a higher-group that can be respected by the ground state is enlarged when killing vectors are available. We will discuss these in more detail in \cref{sec:phases}. Furthermore, when the theory under consideration is embedded on a brane in a higher-dimensional spacetime, as in \cref{sec:embedding}, the collection $\phi$ also involves the embedding functions $X^\mu(\sigma)$ modded by diffeomorphisms on the brane.

Our guiding principle will be the invariance of the partition function \eqref{partitionf} under time-independent background gauge transformations \eqref{2l+2group}, satisfying
\begin{equation}
    \lie_\beta \Lambda_q = \lie_\beta \Lambda_{2q+1} = 0~~.
\end{equation}
Borrowing the philosophy from hydrodynamics, the effective action can be arranged in a derivative expansion and, as usual, we take the Goldstone fields to be of order $\mathcal{O}(\partial^{-1})$ in derivatives. Once we specify the general form of the equilibrium action, we will proceed to derive the equilibrium constitutive relations.  In particular, we find the equilibrium expressions for the higher-form currents and energy-momentum tensor in terms of the appropriate thermodynamic variables, at ideal order in derivatives. 
 
In thermal equilibrium, the fluid velocity can always be aligned with a timelike Killing vector, and we can express it as $u^\mu = T\beta^\mu = \delta^\mu_t/\sqrt{-g_{tt}}$, normalised as $u^\mu u_\mu = -1$, where $T = T_0/\sqrt{-g_{tt}}$ is the local redshifted temperature. This can be used to decompose an arbitrary $p$-form $\xi$ into time and purely spatial components, namely
\begin{equation}
    \xi=-u\wedge i_{u}\xi+\xi_{\Sigma}~~,
    \label{eq:spatial-decomposition}
\end{equation}
where $i_u$ denotes the interior product with respect to $u^\mu$. 
It it also useful to note the identity $*\xi=\star (u\wedge \xi)$, where $*$ is the spatial Hodge duality operation defined with respect to the (induced) metric on the Cauchy slice $\Sigma$. The spatial integrals can be defined as $\int_\Sigma X = - \int_M u\wedge X$ for some $(d-1)$-rank differential form $X$. Taking $X= *f$ for some scalar $f$, we have $\int_\Sigma {*f} = \int_M \star f$. 

\subsection{Equilibrium phases}
\label{sec:phases}

In this subsection, we consider different equilibrium phases of systems with $(2q+2)$-group symmetry, featuring different patterns of spontaneous symmetry breaking. Systems with ordinary 0-form symmetries typically exhibit two phases at finite temperature: spontaneously unbroken and spontaneously broken. On the other hand, the symmetry parameter of a $p$-form symmetry carries spacetime indices and, therefore, can also exhibit an intermediate ``temporal spontaneous symmetry breaking'' pattern where the $p$-form symmetry is only spontaneously broken in the time direction~\cite{Armas:2018atq, Armas:2018zbe}. This gives rise to a $(p-1)$-form gapless mode in the low energy description, regarded as a temporal Goldstone field. The temporal Goldstone is just the time-component of the full $p$-form Goldstone, when the latter is available. 

We find that a system with $U(1)^{(q)}\times_{\kappa} U(1)^{(2q+1)}$ symmetry in $d>2q+2$ spacetime dimensions can admit 3 non-trivial equilibrium phases at finite temperature, given by the table below
\begin{center}
\def\arraystretch{1.3}
\begin{tabular}{c|cc}
    $(d>2q+2)$ & $U(1)^{(q)}$-temporally broken &  
    $U(1)^{(q)}$-completely broken \\
    \hline
    $U(1)^{(2q+1)}$-temporally broken & T-T & C-T  \\
    $U(1)^{(2q+1)}$-completely broken & & C-C
\end{tabular}
\end{center}
Analogously to our discussion in \cref{sub-critical}, the $U(1)^{(2q+1)}$ part of the higher-group exhibits quasi-long-range temporal/complete order in $d=2q+3$ dimensions instead of true spontaneous symmetry breaking. In the remainder of this section, quasi-long-range order is understood where applicable.
Note that there are no phases in the bottom-left corner of this table because when the $U(1)^{(2q+1)}$ part of the symmetry is  temporally/completely spontaneously broken, the $(2q+2)$-group structure also forces the $U(1)^{(q)}$ part of the symmetry to be temporally/completely spontaneously broken.
Phases with spontaneously unbroken $U(1)^{(2q+1)}$ symmetry in $d>2q+3$, or without quasi-long-range order in $d=2q+3$, while allowed by symmetries, do not carry a non-zero thermodynamic $(2q+1)$-form density. Hence we do not consider these in this work. See~\cite{Armas:2018zbe} for related discussion for 1-form symmetries. 

This situation is qualitatively different in sub-critical dimensions $d=2q+2$, where there is no spontaneous symmetry breaking or quasi-long-range order in the $U(1)^{(2q+1)}$ sector. In this case, we only have 2 phases
\begin{center}
\def\arraystretch{1.3}
\begin{tabular}{c|cc}
    $(d=2q+2)$ & $U(1)^{(q)}$-temporally broken &  
    $U(1)^{(q)}$-completely broken \\
    \hline
    $U(1)^{(2q+1)}$-unbroken & T-U & C-U
\end{tabular}
\end{center}
with constant $U(1)^{(2q+1)}$ current $\star\cJ_{2q+2} = Q_{(2q+1)}$. As we discussed in \cref{sec:embedding}, these phases are particularly interesting when the theory is embedded on a dynamical worldvolume in a higher-dimensional spacetime. 

In the following, we will consider all these phases individually from the least to most spontaneously broken symmetries. We will derive the constitutive relations for higher-form currents and energy-momentum tensor in each of the phases, together with the configuration equations for the associated (temporal) Goldstone fields.

\subsubsection{T-U phase (sub-critical dimensions)}
\label{sec:TU}

We start with the T-U phase by considering the spontaneous breaking of the temporal part of $U(1)^{(q)}$ symmetry, leaving the rest of the symmetries intact. To this end, we introduce a temporal Goldstone field $\varphi_{q-1}$, satisfying $i_\beta \varphi_{q-1} = 0$, with transformation
\begin{equation}\label{eq:temp-2}
    \delta \varphi_{q-1}=-i_{\beta}\Lambda_{q}~~.
\end{equation}
It follows that we can define the invariant chemical potential as 
\begin{equation}\label{eq:mu-2}
    \frac{\mu_{q}}{T} =-\df\varphi_{q-1}+i_{\beta}A_{q+1}~~,
\end{equation}
which obeys the Bianchi identity 
\begin{equation}\label{eq:mu2-Bianchi}
    \df\!\lb\frac{\mu_{q}}{T}\rb = -i_\beta G_{q+2}~~,
\end{equation} 
and the spatiality condition $i_\beta \mu_q = 0$. 
In order for this phase to possess a non-trivial $(2q+1)$-form sector, we need an invariant that contains $B_{2q+2}$, so that the action is capable of reproducing a $\mathcal{J}_{2q+2}$ current. Unfortunately, without introducing spontaneous symmetry breaking in the $U(1)^{(2q+1)}$ sector, the best we can do is work in sub-critical dimensions $d=2q+2$, where we can construct an expression that is invariant up to a total derivative. The action takes the generic form
\begin{equation}\label{eq:action-TU}
    S_{E} = \int_\Sigma {*\hat P(T,\mu_q)} 
    + Q_{(2q+1)}\int_{\Sigma}\psi_{2q+1}~~,
\end{equation}
where $\hat P$ is a function of $T$ and $\mu_q$. For instance, we can write down a quadratic term in the effective action via ${*\hat P} = \half \chi \,\mu_{q}\wedge \ast \mu_{q}$,
where $\chi$ is the $U(1)^{(q)}$ susceptibility. Above, we have also defined the spatial top-form
\begin{align}\label{eq:psi-def}
    \frac{1}{T}\psi_{2q+1}
    &= i_{\beta}B_{2q+2}
    + 2\kappa\, \df\varphi_{q-1}\wedge A_{q+1,\Sigma}
    - \kappa\, i_{\beta}A_{q+1}\wedge A_{q+1,\Sigma}
    - \kappa\,\df(Tu) \wedge \varphi_{q-1} \wedge \df\varphi_{q-1}~~,
\end{align}
where we have used the notation \eqref{eq:spatial-decomposition} for the spatial part of differential forms.
The transformation of $\psi_{2q+1}$ is
\begin{align}
    \frac1T \delta \psi_{2q+1}
    &= i_{\beta}\df\Lambda_{2q+1}
    + 2\kappa\, \df\varphi_{q-1}\wedge \df\Lambda_{q}
    + \kappa\,\df(Tu \wedge i_\beta\Lambda_{q})\wedge \df\varphi_{q-1}
    + \kappa\,\df(Tu \wedge \varphi_{q-1})\wedge \df i_\beta\Lambda_q \nn\\
    &= \df\Big(
    - i_{\beta}\Lambda_{2q+1}
    + \kappa\, \varphi_{q-1}\wedge (\df \Lambda_{q})_{\Sigma}
    + \kappa\,\Lambda_{q,\Sigma} \wedge\df\varphi_{q-1}
    \Big)~,
    \label{eq:psi-transform}
\end{align}
resulting in a boundary term in the action.\footnote{Note that $\int_{\Sigma} T \df X = -\int_{M} Tu\wedge \df X = \int_{M} \df(Tu\wedge X) -\int_M \df(Tu)\wedge X$ for any $(d-2)$-rank form $X$. Given that $X$ is purely spatial, the last term vanishes in equilibrium because $\df(Tu)\wedge X = -Tu\wedge i_\beta \df(Tu)\wedge X = -Tu\wedge \lie_\beta(Tu)\wedge X = 0$ when $\beta^\mu$ is a Killing vector.}

Drawing inspiration from the zero temperature case, we may introduce a non-dynamical Stueckelberg field $\varphi_{2q}$ transforming in the same way as the term inside the differential in \cref{eq:psi-transform}, allowing us to replace $\psi_{2q+1}$ with an exact invariant $\mu_{2q+1} = \psi_{2q+1} - T\df\varphi_{2q}$. In fact, in higher dimensions, $\varphi_{2q}$ and $\mu_{2q+1}$ are precisely the higher-rank temporal Goldstone field and chemical potential respectively, that we will encounter in \cref{sec:TT}. We can then recast the action \cref{eq:action-TU} as
\begin{equation}\label{eq:action-TT}
    S_{E} = \int_\Sigma {*P(T,\mu_q,\mu_{2q+1})}~~,
\end{equation}
where $P = \hat P + Q_{(2q+1)}{*\mu_{2q+1}}$.
Note that the dependence on $\varphi_{2q}$ is a total-derivative and identically drops out form the integral.
In what follows, we compute the configuration equations and constitutive relations arising from the action in \eqref{eq:action-TT} in $d=2q+2$ spacetime dimensions, with a strictly linear coupling of $\mu_{2q+1}$, while the dependence on $\mu_{q}$ is left arbitrary.

The function $P$ in the action \eqref{eq:action-TT} can be identified as the thermodynamic pressure of the fluid. Its variation with respect to the thermodynamic variables can be used to define other thermodynamic quantities as\footnote{We use the ``dot-product'' notation for two $p$-forms $A$ and $B$ as $A\cdot B = \frac{1}{p!} A^{\mu\ldots}B_{\mu\ldots}$.}
\begin{align}\label{eq:EOS-TT}
     \delta P
     &= s\,\delta T
     + n_q\cdot \delta \mu_q
     + n_{2q+1}\cdot \delta\mu_{2q+1}
     - \frac{1}{2}r^{\mu \nu}\delta g_{\mu \nu}~~, \nn\\
     \epsilon 
     &= Ts+\mu_q\cdot n_q+\mu_{2q+1}\cdot n_{2q+1} - P~~,
\end{align}
where $n_{2q+1}=Q_{(2q+1)}{\ast 1}$. Here $s$ is the entropy density, $n_q$ the $q$-form density, $n_{2q+1}$ the $(2q+1)$-form density, $r^{\mu\nu}$ the anisotropic stress tensor, and $\epsilon$ the energy density. The $r^{\mu\nu}$ contribution is required because $P$ is a scalar, and is fixed by local Lorentz invariance to be
\begin{equation}
     r^{\mu \nu}
     = \frac{1}{(q-1)!}
     (n_q)^{\mu}_{~\rho\ldots}
     (\mu_q)^{\nu\rho\ldots}
     + \frac{1}{(2q)!}
     (n_{2q+1})^{\mu}_{~\rho\ldots}
     (\mu_{2q+1})^{\nu\rho\ldots}~~,
\end{equation}
together with the constraint  $r^{[\mu \nu]}=0$.
Varying the equilibrium effective action with respect to the higher-form background gauge fields, we find the higher-form currents
\begin{align}\label{constitemporal22-J}
    J_{q+1}
    &=u\wedge n_{q} 
    - (-)^d 2\kappa\, {\ast\!\lb\mu_{q} \wedge {\ast n_{2q+1}}\rb}~~, \nn\\
    \mathcal{J}_{2q+2}
    &= u \wedge n_{2q+1}~.
\end{align}
The second term in $J_{q+1}$ is a purely higher-group effect. In sub-critical dimensions, this term is simply $- 2\kappa Q_{(2q+2)} {\ast\mu_{q}}$, but the more general form will be helpful later in \cref{sec:TT}. At non-zero $q$-form chemical potential and $(2q+1)$-form density, this term survives even when the background fields vanish.
The same effect was observed in \cite{Iqbal:2020lrt} for the 2-group. The energy-momentum tensor for this phase reads
\begin{equation}\label{eq:EM-tensor}
     T^{\mu \nu}=(\epsilon+P)u^{\mu}u^{\nu}+Pg^{\mu \nu}
     - r^{\mu \nu}
     + 2\upsilon^{(\mu} u^{\nu)}
     ~~,
\end{equation}
where we have identified the 1-form heat-flux contribution
\begin{equation}
    *\upsilon = -\kappa\, \mu_q\wedge\mu_q\wedge {*n_{2q+1}}~~,
\end{equation}
which is also characteristic of the higher-group structure in thermal equilibrium. This contribution is also responsible for violating the Landau frame condition at ideal order, i.e. $T^{\mu\nu} u_\nu \neq -\epsilon\, u^\mu$, where we recall that the fluid velocity is aligned along the equilibrium thermal vector $\beta^\mu = \delta^\mu_t/T_0$. Finally, the temporal Goldstone $\varphi_q$ obeys the configuration equation
\begin{equation}\label{config1}
    \df{\ast n_{q}} = 2\kappa\lb 
    G_{q+2,\Sigma}
    + \df (Tu) \wedge \frac{\mu_q}{T} \rb
    \wedge {* n_{2q+1}}~.
\end{equation}
It is straightforward to check that the above configuration equation combined with the constitutive relations reproduces the $(2q+2)$-group conservation laws.

\subsubsection{C-U phase (sub-critical dimensions)}
\label{sec:CU}

We now consider the C-U phase where the $U(1)^{(q)}$ symmetry is completely spontaneously broken, while the $U(1)^{(2q+1)}$ symmetry still remains unbroken. This means that we have access to the full Goldstone $\phi_{q}$ with transformation $\delta \phi_{q}=-\Lambda_{q}$. We have set the Goldstone charge $c_\phi = 1$ for simplicity.
The temporal Goldstone $\varphi_{q-1}$ is related to $\phi_{q}$ via $\varphi_{q-1}=i_{\beta} \phi_{q}$. We can use this to define the chemical potential $\mu_q$ which is the same as in \cref{eq:mu-2}. Note the simple relation between $q$-form chemical potential and $(q+1)$-form field strength
\begin{equation}
    \iota_\beta F_{q+1} = \frac{\mu_q}{T}~~.
\end{equation}
For the record, let us also write down the spatial components of the respective Bianchi identity
\begin{equation}\label{eq:np-Bianchi}
    \df{*\tilde n_{p}}
    = 
    (-)^d\df(Tu)\wedge \frac{\mu_q}{T}
    + (-)^d G_{q+2,\Sigma}~~,
\end{equation}
where $p=d-2-q$ and we have defined $\tilde n_p = *F_{q+1}$. Together with \cref{eq:mu2-Bianchi}, this gives rise to the full Bianchi identity \eqref{Bianchi3}. The introduction of the invariant thermodynamic variable $\tilde n_p$ is motivated from the fact that the Bianchi identities can themselves be interpreted as emergent higher-form conservation laws. From this perspective, which is elaborated in \cref{sec:dual},  $\tilde n_p$ becomes a conserved $p$-form density in its own right.
As in \cref{sec:TU}, the only quantity at our disposal on which the action can depend containing the background field $B_{2q+2}$ is $\psi_{2q+1}$ defined in \cref{eq:psi-def}. However, $\psi_{2q+1}$ is only symmetry-invariant up to a total-derivative term, forcing us to sub-critical dimensions $d=2q+2$ and a linear to $\psi_{2q+1}$ analogous to \cref{eq:action-TU}. What distinguishes this phase from its T-U counterpart is the fact that the equilibrium action can now also depend on the spatial components of $F_{q+1}$.  For example, we may write down a term in the equilibrium action $-\int_\Sigma \half f F_{q+1}\wedge *F_{q+1}$.
More generally, the equilibrium effective action for this phase takes the form
\begin{equation}\label{eq:action-CU}
    S_{E} = \int_{\Sigma} {*\hat P}(T,\mu_q,\tilde n_p)
    + Q_{(2q+2)}\int_{\Sigma} \psi_{2q+1}~~.
\end{equation}
Following our discussion in \cref{sec:TU}, we can make our life simpler by introducing a non-dynamical Stueckelberg field $\varphi_{2q}$ used to define the invariant chemical potential $\mu_{2q+1}$, enabling us to recast the action as
\begin{equation}\label{eq:action-CT}
    S_{E} = \int_\Sigma {*P(T,\mu_q,\mu_{2q+1},\tilde n_p)}~~,
\end{equation}
where the thermodynamic pressure is defined as $P = \hat P + Q_{(2q+1)}{*\mu_{2q+1}}$. The thermodynamic relations in this phase are given as
\begin{align}\label{eq:EOS-CT}
    \delta P
    &= s\,\delta T
    + n_q\cdot \delta \mu_q
    + n_{2q+1}\cdot \delta\mu_{2q+1}
    - \ast \Tilde{\mu}_{p}\cdot\delta{\ast\tilde{n}_p}
    - \frac{1}{2}r^{\mu \nu}\delta g_{\mu \nu}~~, \nn\\
    \epsilon 
     &= Ts+\mu_q\cdot n_q+\mu_{2q+1}\cdot n_{2q+1} - P~~,
\end{align}
where $n_{2q+1}=Q_{(2q+1)}{\ast 1}$ and the anisotropic stress tensor is given by
\begin{align}
     r^{\mu \nu}
     &= \frac{1}{(q-1)!}
     (n_q)^{\mu}_{~\rho\ldots}
     (\mu_q)^{\nu\rho\ldots}
     + \frac{1}{(2q)!}
     (n_{2q+1})^{\mu}_{~\rho\ldots}
     (\mu_{2q+1})^{\nu\rho\ldots} \nn\\
     &\qquad 
     - \frac{1}{q!}
     (*n_p)^{\mu}_{~\rho\ldots}
     (*\mu_p)^{\nu\rho\ldots}~~.
\end{align}
This variation introduces a new thermodynamic quantity $*\tilde\mu_p$ which can be identified as the dissipationless equilibrium $U(1)^{(q)}$ superflow. To wit, varying the general equilibrium action, we obtain the higher-form currents
\begin{align}\label{eq:consti-CU}
    J_{q+1}
    &=u\wedge n_{q} 
    - (-)^d{*\tilde\mu_p}
    - (-)^d 2\kappa\, {\ast\!\lb\mu_{q} \wedge {\ast n_{2q+1}}\rb}~~, \nn\\
    \mathcal{J}_{2q+2}
    &= u \wedge n_{2q+1}~.
\end{align} 
The energy-momentum tensor takes the same form as \cref{eq:EM-tensor}, except that the presence of $\tilde{n}_{p}$ results in a new term inside the heat-flux contribution, namely
\begin{align}
     *\upsilon
     =
     - \mu_q\wedge \tilde{\mu}_p
     - \kappa\,\mu_q\wedge\mu_q\wedge {*n_{2q+1}}~~.
\end{align}
The configuration equations for the temporal and spatial components of the Goldstone field $\phi_q$ are given as
\begin{align}\label{configspatial2}
    \df{\ast n_{q}} 
    &= \df(T u) \wedge \frac{\tilde\mu_p}{T}
    + 2\kappa\lb 
    G_{q+2,\Sigma}
    + \df (Tu) \wedge \frac{\mu_q}{T} \rb
\wedge {* n_{2q+1}}~, \nn\\
    \df\frac{\tilde{\mu}_p}{T} 
    &= 0~.
\end{align}
Note that the configuration equation for the temporal Goldstone has changed compared to \cref{config1} in the T-U phase because of $\tilde\mu_p$.

\subsubsection{T-T phase}
\label{sec:TT}

We now turn to the T-T phase for which the temporal part of the $U(1)^{(2q+1)}$ symmetry is spontaneously broken in $d>2q+2$ dimensions. To this end, we introduce a $(2q)$-form temporal Goldstone field $\varphi_{2q}$, satisfying $i_\beta \varphi_{2q} = 0$, with transformation law
\begin{equation}\label{temporalgold4}
    \delta \varphi_{2q}
    = - i_{\beta}\Lambda_{2q+1}
    + \kappa\, \varphi_{q-1}\wedge (\df \Lambda_{q})_{\Sigma}
    + \kappa\,\Lambda_{q,\Sigma} \wedge\df\varphi_{q-1}~,
\end{equation}
where the right-hand side is precisely the quantity inside the differential in \cref{eq:psi-transform}.
Observe that, due to the higher-group structure, this transformation rule contains additional terms compared to the simple transformation rule of $\varphi_{q-1}$ in \cref{eq:temp-2}.\footnote{Had we used the background gauge transformations in the form of \cref{B2many}, with the higher-rank gauge parameter $\tilde\Lambda_{2q+1} = \Lambda_{2q+1} + \kappa\,\Lambda_q \wedge A_{q+1}$ we would instead define the temporal-Goldstone $\tilde\varphi_{2q} = \varphi_{2q} - \kappa\, \varphi_{q-1} \wedge A_{q+1,\Sigma}$ with the following transformation law
\begin{equation}
    \delta \tilde\varphi_{2q}
    = - i_{\beta}\tilde\Lambda_{2q+1}
    - \kappa\,\Lambda_{q,\Sigma} \wedge \frac{\mu_q}{T}~.
\end{equation}
This is the $(2q+2)$-group analogue of the transformation introduced in \cite{Iqbal:2020lrt} for a 2-group. However, note that this transformation depends on the background field $A_{q+1}$ implicitly via $\mu_q$.
} Using \eqref{temporalgold4}, we can define the invariant $(2q+1)$-form chemical potential
\begin{align}\label{eq:mu-5}
   \frac{\mu_{2q+1}}{T}
   &= i_{\beta}B_{2q+2}
   - \df\varphi_{2q} \nn\\
   &\qquad 
    + 2\kappa\, \df\varphi_{q-1}\wedge A_{q+1,\Sigma}
    - \kappa\, i_{\beta}A_{q+1}\wedge A_{q+1,\Sigma}
    - \kappa\,\df(Tu) \wedge \varphi_{q-1} \wedge \df\varphi_{q-1}~~,
\end{align}
satisfying the Bianchi identity  
\begin{equation}\label{eq:mu5-Bianchi}
    \df\!\lb\frac{\mu_{2q+1}}{T} \rb
    = -i_\beta G_{2q+3}
    - \kappa\lb 2G_{q+2,\Sigma}
    + \df(Tu)\wedge \frac{\mu_q}{T}\rb \wedge \frac{\mu_q}{T}~~,
\end{equation}
and the spatiality condition $i_\beta \mu_{2q+1} = 0$. In parallel to the zero temperature (and in principle time-dependent) case, we observe the necessity of introducing the $U(1)^{(q)}$ temporal Goldstone $\varphi_{q-1}$ in order to define the invariant $U(1)^{(2q+1)}$ chemical potential $\mu_{2q+1}$. This points towards the fact that in order to spontaneously break the temporal part of the $U(1)^{(2q+1)}$ symmetry, we need to spontaneously break the temporal part of the $U(1)^{(q)}$ symmetry as well. 

The effective action for the T-T phase is written in the same way as in \cref{eq:action-TT}, but with arbitrary non-linear dependence on the higher-rank chemical potential $\mu_{2q+1}$. 
As an example, at quadratic level, we may write down $*P = \half\chi\mu_q\wedge \ast \mu_q+\half\chi'\mu_{2q+1}\wedge \ast \mu_{2q+1}$. The remainder of the discussion for constitutive relations is exactly the same as in the T-U phase following \cref{eq:EOS-TT}, but with general higher-rank density $n_{2q+1}$. The configuration equation for the temporal Goldstone $\varphi_{2q}$ is simply
\begin{equation}\label{config2}
    \df{\ast n_{2q+1}}=0~~,
\end{equation}
which used to be trivial in the T-U phase, while the configuration equation for the temporal-Goldstone $\varphi_{q-1}$ is still given by \cref{config1}.

\subsubsection{C-T phase}
\label{sec:CT}

The C-T phase features complete spontaneous breaking in the $U(1)^{(q)}$ sector and temporal spontaneous breaking in the $U(1)^{(2q+1)}$ sector in dimensions $d>2q+2$, introducing the full Goldstone $\phi_q$ and the temporal Goldstone $\varphi_{2q}$. 
Reversing this logic, namely temporally spontaneously breaking the $U(1)^{(q)}$ symmetry and completely spontaneously breaking the $U(1)^{(2q+1)}$ symmetry, is forbidden by the higher-group structure because the residual symmetry generators do not form a proper subgroup of the $(2q+2)$-group.
The general effective action for the C-T phase is still given by \cref{eq:action-CU}, but with a general, not necessarily linear dependence on $\mu_{2q+1}$. For example, one can write down the action
 \begin{equation}
      *P = \frac{\chi}{2}\mu_{q}\wedge \ast \mu_{q} -\frac{f}{2}F_{q+1} \wedge \ast F_{q+1} +\frac{\chi'}{2}\mu_{2q+1}\wedge \ast \mu_{2q+1}~~.
\end{equation}
The remainder of the discussion for constitutive relations is exactly the same as in the C-U phase following \cref{eq:EOS-CT}, but with general higher-rank density $n_{2q+1}$. The configuration equations for the temporal and spatial components of $\phi_{q}$ are still given by \cref{configspatial2}, while that for the temporal-Goldstone $\varphi_{2q}$ is simply given by \cref{config2}.

\subsubsection{C-C phase}

We now turn to the final phase for which the $U(1)^{(2q+1)}$ symmetry is completely-spontaneously broken in dimensions $d>2q+2$, meaning that we bring in the full Goldstone $\Phi_{2q+1}$ with transformation laws \eqref{Golstonetrans5}. We choose also here the normalization $c_\Phi = 1$. In order to write an invariant field strength for this field, as we already witnessed in our analysis at zero temperature, we need to completely spontaneously break the $U(1)^{(q)}$ symmetry as well. Since complete spontaneous breaking by definition includes partial spontaneous breaking, this phase features the most general symmetry breaking pattern that we will encounter in this paper. The temporal Goldstone $\varphi_{2q}$ can be obtained in terms of $\Phi_{2q+1}$ as
\begin{equation}
    \varphi_{2q}
    = i_{\beta}\Phi_{2q+1}
    - \kappa\,\varphi_{q-1}\wedge (\df \phi_{q})_\Sigma~~,
\end{equation}
defined to be purely spatial, transforming as \cref{temporalgold4}. In this phase we have at our disposal the full field strength $\mathcal{F}_{2q+2}$ defined \cref{invariant2}, which satisfies the relation
\begin{equation}
    i_{\beta}\mathcal{F}_{2q+2}
    = \frac{\mu_{2q+1}}{T}
    + \kappa \frac{\mu_{q}}{T}\wedge F_{q+1,\Sigma}~~.
\end{equation}
The spatial components of the associated Bianchi identity is given as
\begin{equation}\label{eq:n-higher-Bianchi}
    \df{*\tilde{n}_{p-q-1}}
    = 
    - \mathcal{G}_{2q+3,\Sigma}
    - \df(Tu)\wedge \frac{\mu_{2q+1}}{T}
    - (-)^d\kappa \lb G_{q+2,\Sigma}
    + \df(Tu)\wedge \frac{\mu_{q}}{T} \rb \wedge {*\tilde n_p}
    ~~,
\end{equation}
where $\tilde{n}_{p-q-1}=-{\ast \mathcal{F}_{2q+2}}$.
This together with \cref{eq:mu5-Bianchi} gives rise to the full Bianchi identity \eqref{Bianchi6}.
To be explicit, we can write down for instance the following quadratic action in the C-C phase
\begin{equation}
     S_{E}=\int_{\Sigma_{d-1}}\frac{\chi}{2}\mu_{q}\wedge \ast \mu_{q} -\frac{f}{2}F_{q+1} \wedge \ast F_{q+1} +\frac{\chi'}{2}\mu_{2q+1}\wedge \ast \mu_{2q+1}-\frac{f'}{2}\mathcal{F}_{2q+2}\wedge \ast \mathcal{F}_{2q+2}~.
\end{equation}
In particular, the effective action can now depend on the additional quantity $\tilde{n}_{p-q-1}=-{\ast \mathcal{F}_{2q+2}}$.
More generally, however, the effective action takes the form
\begin{equation}
    S_E = \int_\Sigma *P(\mu_{q},\mu_{2q+1},\Tilde{n}_{p}, \Tilde{n}_{p-q-1})~~.
\end{equation}
We should emphasize that there is no equivalent of this phase in sub-critical dimensions $d=2q+2$, simply because $\cF_{2q+2}$ is a spacetime top-form and its spatial Hodge-dual ${\ast \mathcal{F}_{2q+2}}$ is identically zero.

The thermodynamics in the C-C phase takes the general form
\begin{align}
     \delta P
     &= s\,\delta T
     + n_{q}\cdot \delta \mu_{q}
     + n_{2q+1}\cdot \delta \mu_{2q+1}
     - \ast \Tilde{\mu}_{p}\cdot\delta{\ast \Tilde{n}_{p}}
     - \ast \Tilde{\mu}_{p-q-1}\cdot\delta{\ast \Tilde{n}_{p-q-1}}
     -\frac{1}{2}r^{\mu \nu}\delta g_{\mu \nu}~~, \nn\\
     \epsilon 
     &= Ts+\mu_q\cdot n_q+\mu_{2q+1}\cdot n_{2q+1} - P~~,
\end{align}
where the anisotropic stress tensor reads
\begin{align}
     r^{\mu \nu}
     &= \frac{1}{(q-1)!}
     (n_q)^{\mu}_{~\rho\ldots}
     (\mu_q)^{\nu\rho\ldots}
     + \frac{1}{(2q)!}
     (n_{2q+1})^{\mu}_{~\rho\ldots}
     (\mu_{2q+1})^{\nu\rho\ldots} \nn\\
     &\qquad
     - \frac{1}{q!}
     (*n_p)^{\mu}_{~\rho\ldots}
     (*\mu_p)^{\nu\rho\ldots}
     -\frac{1}{(2q+1)!}(\ast \tilde{n}_{p-q-1})^{\mu}_{~\rho\dots}(\ast \tilde{\mu}_{p-q-1})^{\nu \rho \ldots}~~.
\end{align}
Varying the effective action, we are led to the higher-form currents
\begin{align}\label{currentJ3COM/COM}
    J_{q+1}
    &= u\wedge n_{q}
    - (-1)^{d}{\ast \Tilde{\mu}_{p}}
    - (-1)^{d}2\kappa\,{\ast(\mu_{q}\wedge \ast n_{2q+1})}
    + \kappa\,{\ast (\ast \tilde{n}_{p}\wedge \tilde{\mu}_{p-q-1})}~~, \nn\\
    \mathcal{J}_{2q+2}
    &= u\wedge n_{2q+1} + \ast \tilde{\mu}_{p-q-1}~~.
\end{align}
The energy-momentum tensor still takes the same form as \cref{eq:EM-tensor}, except that the full heat-flux contribution is given as
\begin{align}\label{Tmncomplete2complete5}
     *\upsilon
     &= -\mu_{q}\wedge \tilde{\mu}_{p}
     - \kappa\, \mu_{q}\wedge\mu_{q}\wedge \ast n_{2q+1} 
     + \mu_{2q+1} \wedge \tilde{\mu}_{p-q-1}
    + \kappa (-1)^{d} \mu_{q}\wedge \ast \tilde{n}_{p} \wedge \tilde{\mu}_{p-q-1}~~.
\end{align}
The configuration equations for the temporal and spatial parts of the Goldstone fields $\phi_{q}$ and $\Phi_{2q+1}$ are given as
\begin{align}\label{eq:config-CC}
    \df{* n_{q}} 
    &= \df(T u) \wedge \frac{\tilde\mu_p}{T}
    + 2\kappa\lb 
    G_{q+2,\Sigma}
    + \df (Tu) \wedge \frac{\mu_q}{T} \rb
    \wedge {* n_{2q+1}}
    - \kappa (-1)^{d}\df(Tu)\wedge \ast \tilde{n}_{p}\wedge \frac{\tilde{\mu}_{p-q-1}}{T}~, \nn\\
    \df\frac{\tilde\mu_p}{T}
    &=- \kappa \left(G_{q+2,\Sigma} + \df(Tu)\wedge \frac{\mu_{q}}{T}\right)\wedge \frac{\tilde{\mu}_{p-q-1}}{T}~~, \nn\\
    \df{* n_{2q+1}}
    &= \df(Tu)\wedge \frac{\tilde{\mu}_{p-q-1}}{T} ~~, \nn\\
    \df\frac{\tilde\mu_{p-q-1}}{T} 
    &=0~~.
\end{align}
One can verify that the constitutive relations along with the configurations equations are consistent with the $(2q+2)$-group conservation laws for this phase as well.

\subsection{Self-dual limit and the thermal M5 brane}
\label{sec:M5-finite-T}

Given the exposition of phases with $(2q+2)$-group symmetry at finite temperature in \cref{sec:phases}, we now analyze the M5 brane at finite temperature, from the perspective of the underlying 6-group symmetry. As we have discussed at length in \cref{sec:M5-brane}, the $U(1)^{(2)}$ part of the 6-group symmetry on the M5 brane is spontaneously broken at zero temperature, giving rise to the 2-form field $\phi_2$ living on its worldvolume, while the dimensionality of the brane is sub-critical with respect to the $U(1)^{(5)}$ part and thus it cannot be spontaneously broken. The only subtlety is that the field strength $F_3$ associated with the Goldstone $\phi_2$ is required to be self-dual on account of supersymmetry. While we were able to construct a zero-temperature effective action for the bosonic M5 brane consistent with self-duality in \cref{sec:M5-brane}, we were unable to provide a symmetry perspective that may directly yield the self-duality constraint either offshell or onshell. Interestingly, this situation is not as complicated at finite temperature.

Let us start with the C-U in phase in sub-critical dimensions discussed in \cref{sec:CU}, where the $U(1)^{(q)}$ symmetry is completely-spontaneously broken and the $U(1)^{(2q+1)}$ symmetry is unbroken, giving rise to the $q$-form Goldstone $\phi_q$ with the temporal-Goldstone $\varphi_{q-1} = i_\beta \phi_q$. Naively, this is the phase that applies to the M5 brane for $q=2$, with the added self-duality constraint. The non-linear self-duality condition is given in \cref{eq:non-linear-self-duality}. Plugging in the C-U phase constitutive relations from \cref{eq:consti-CU} in sub-critical dimensions $d=2q+2$, i.e. $p=q$, the time and space components of this equation are given as
\begin{subequations}
\begin{alignat}{3}
    \tilde n_p &= \frac{1}{2\kappa Q_{(2q+1)}} n_{q} \label{eq:self-duality-finiteT-1}
    &\quad\implies\quad&&
    \tilde n_p &= \frac{1}{2\kappa Q_{(2q+1)}} \frac{\delta S_E}{\delta \mu_q}
    \bigg|_{T,\tilde n_p,\mu_{2q+1}}~~, \\
    \tilde\mu_p &= 0
    &\quad\implies\quad&&
    0 &= \frac{\delta S_E}{\delta\tilde n_p}
    \bigg|_{T,\mu_q,\mu_{2q+1}}~~. \label{eq:self-duality-finiteT-2}
\end{alignat}
\end{subequations}
In particular, note that by switching off the dependence on $\tilde n_p$, the C-U phase equilibrium action in \cref{eq:action-CT} reduces to that of the T-U phase in \cref{eq:action-TT}. In other words, the $(2q+2)$-group C-U phase with self-duality is effectively described by the T-U at finite temperature, where the $U(1)^{(q)}$ part of the $(2q+2)$-group symmetry is only spontaneously broken along the time-direction. For our case of interest, this means that the bosonic M5 brane at finite temperature is actually described by the T-U phase.

The temporal spontaneous symmetry breaking patter in the T-U phase automatically takes care of the self-duality constraint. Using the action \eqref{eq:action-TU} in the non-trivial part of the self-duality condition \cref{eq:self-duality-finiteT-1}, we can find
\begin{equation}
    *F_{q+1} 
    = \frac{1}{2\kappa Q_{(2q+1)}} 
    *\!\bfrac{\delta{*\hat P(T,\mu_q)}}{\delta\mu_q}~~, \qquad 
    \iota_u F_{q+1} = \mu_{q}~~.
    \label{eq:finite-T-selfduality}
\end{equation}
Since the right-hand side of the first equation only depends on the ``electric components'' $i_u F_{q+1}$ and not on the ``magnetic components'' $*F_{q+1}$, this equation can effectively be seen as defining the magnetic components of $F_{q+1}$ in terms of the electric components.  Every T-U phase effective action of the form in \cref{eq:action-TU} describes a self-dual theory, without the need for imposing additional constraints. The reason for this conceptual simplification can be traced back to the broken Lorentz invariance at finite temperature. The primary complication while writing down self-dual effective actions at zero temperature arises from simultaneously requiring manifest Lorentz-invariance. This fact was exploited in~\cite{Perry:1996mk}, where the authors constructed a zero temperature effective action for the bosonic M5 brane by giving up manifest Lorentz-invariance. Nonetheless, the authors illustrated that the equations of motion arising from this effective action were indeed Lorentz invariant. By contrast, Lorentz invariance at finite temperature is broken by the thermal frame of reference or the fluid velocity, circumventing the entire issue.

To summarize, the bosonic M5 brane at finite temperature is described by the T-U phase 6-group invariant equilibrium effective action 
\begin{equation}\label{eq:M5-action-finiteT}
    S_{E} = \int_\Sigma {*\hat P(T,\mu_2)} 
    + Q_{(5)}\int_{\Sigma}\mu_{5}~~,
\end{equation}
obtained from \cref{eq:action-TU} by setting $q=2$. For completeness, let us write also here the definitions of the chemical potentials $\mu_2$ and $\mu_5$, i.e.
\begin{align}
    \frac{\mu_2}{T} 
    &= -\df\varphi_1 + i_\beta A_3~~, \nn\\ 
    \frac{\mu_{5}}{T}
   &= - \df\varphi_{4} 
   + i_{\beta}B_{6}
    - \df\varphi_{1}\wedge A_{3,\Sigma}
    + \half i_{\beta}A_{3}\wedge A_{3,\Sigma}
    + \half \df(Tu) \wedge \varphi_{1} \wedge \df\varphi_{1}~~,
\end{align}
where we have set $\kappa = -1/2$ specific to the M5 brane.
Note that the definition of $\mu_5$ contains a contribution from the Stueckelberg field $\varphi_4$ to make it manifestly gauge-invariant, but it trivially drops out from the effective action as a boundary term. The full field strength $F_{q+1}$ is defined in terms of these as
\begin{equation}
    *F_{3} 
    = \frac{-1}{Q_{(5)}} 
    *\!\bfrac{\delta{*\hat P(T,\mu_2)}}{\delta\mu_2}~~, \qquad 
    \iota_u F_{3} = \mu_{2}~~,
    \label{eq:finite-T-selfduality-M5}
\end{equation}
which identically gives rise the self-duality condition that fixes the magnetic components $*F_3$ in terms of the electric components $\iota_u F_3$. In particular, neither the finite temperature effective action \eqref{eq:M5-action-finiteT} nor the associated self-duality relation \eqref{eq:finite-T-selfduality-M5} are Lorentz-invariant. We also record the constitutive relations specific to the thermal M5 brane, i.e.
\begin{align}\label{eq:consti-thermal-M5}
    J_3 
    &= u\wedge n_2
    + Q_{(5)} {\ast\mu_2}~~, \nn\\
    \cJ_6 
    &= Q_{(5)} u \wedge {*1}~~, \nn\\
    T^{\mu\nu}
    &= (\epsilon+\hat P)u^{\mu}u^{\nu} + \hat Pg^{\mu \nu}
     - n_2{}^{\mu}_{~\rho}\, \mu_2^{\nu\rho}
     + Q_{(5)} {*(\mu_2\wedge\mu_2)}^{(\mu} u^{\nu)}~~,
\end{align}
taken directly from \cref{sec:TU} for $q=2$. The thermodynamic relations are given as
\begin{align}
    \delta \hat P
     &= s\,\delta T
     + \half n_2^{\mu\nu} \delta \mu_{2\mu\nu}
     - \frac{1}{2} n_2{}^{\mu}_{~\rho}\, \mu_2^{\nu\rho} \delta g_{\mu \nu}~~, \nn\\
     \epsilon 
     &= Ts + \half \mu_{2\mu\nu} n_2^{\mu\nu} - \hat P~~.
\end{align}

These results can be compared with the work of~\cite{Armas:2019asf}, where the authors derived the equation of state for thermal M5 branes arising from 11d supergravity.\footnote{See also \cite{Armas:2013ota, Niarchos:2012pn, Niarchos:2012cy, Niarchos:2013ia} for earlier work on thermal M5 branes.} The authors focused on M2 brane configurations with $\mu_2 = \Phi_{(2)}{*(v\wedge w\wedge z)}$, where $\Phi_{(2)}$ is a scalar (not to be confused with our Goldstone field) and $v^\mu$, $w^\mu$, $z^\mu$ are mutually orthogonal normalized spatial vectors. Note that this parametrisation of $\mu_2$ is non-generic; in particular, it implies $\mu_2\wedge\mu_2=0$ and $\tr(\mu_2^{2n}) = 2^{1-n}\tr(\mu^2_2)^n$. This sets the $Q_{(5)}$-dependent vector heat-flux contribution in the energy-momentum tensor in \cref{eq:consti-thermal-M5} to zero. Another simplification that happens because of this is that the equation of state $\hat P(T,\mu_2) = \hat p(T,\Phi_{(2)})$ only depends on $\Phi_{(2)}$ and not on the full $\mu_2$, implying that $n_2 = Q_{(2)}{*(v\wedge w\wedge z)}$ where $Q_{(2)} = \dow\hat p/\dow\Phi_{(2)}$. With these identifications in place, one can verify that the constitutive relations \eqref{eq:consti-thermal-M5} exactly match with those derived in \cite{Armas:2019asf}, with the equation of state given as
\begin{equation}
    \hat p(T,\Phi_{(2)}) = - \frac13 Q_{(5)} 
        \frac{1 + 3(\tanh^2\alpha-\Phi_{(2)}^2)\cosh^2\alpha}{\cosh^2\alpha \sqrt{\tanh^2\alpha-\Phi_{(2)}^2}}~~,
\end{equation}
where the parameter $\alpha$ is determined via
\begin{equation}
    \frac{Q_{(5)} \cosh\alpha}{\sqrt{\tanh^2\alpha-\Phi_{(2)}^2}}
    = \frac{3\Omega_4}{16\pi G_N} \lb\frac{3}{4\pi T}\rb^3~~,
\end{equation}
with $\Omega_4 = 4\pi^2/3$ denoting the volume of a unit 4-sphere and $G_N$ the 11d Newton's constant. Note that because of the simple parametrisation of $\mu_2$, the analysis in~\cite{Armas:2019asf} cannot account for more general dependence of the equation of state on non-linear scalars of the kind $\tr(\mu_2^{2n}) - 2^{1-n}\tr(\mu^2_2)^n$ for $n\geq 2$, which would require one to find 11d supergravity solutions with more non-trivial configurations of M2-M5 branes. This would be an interesting avenue to explore in the future.

\section{Dual formulation and anomalous higher-group symmetries}
\label{sec:dual}

Many physical systems with spontaneously broken continuous global symmetries admit dual descriptions in terms of higher-form symmetries, where the Bianchi identities associated with the Goldstone fields are interpreted as higher-form conservation laws in their own right. The best explored examples of these dualities are: viscoelastic hydrodynamics with spontaneously broken translation symmetry recast in terms of $(d-1)$-copies of $U(1)^{(d-1)}$ symmetries~\cite{Grozdanov:2018ewh, Armas:2019sbe}; $q$-form superfluids with spontaneously broken $U(1)^{(q)}$ global symmetry in terms of $U(1)^{(q)}\times U(1)^{(d-1-q)}$ anomalous symmetry~\cite{Delacretaz:2019brr, Armas:2018zbe, Armas:2023tyx}; and free $U(1)^{(q)}$ gauge theory, including free electromagnetism for $q=0$, in terms of $U(1)^{(q+1)}\times U(1)^{(d-2-q)}$ anomalous symmetry~\cite{Gaiotto:2014kfa}; magnetohydrodynamics in terms of $U(1)^{(1)}$ symmetry~\cite{Grozdanov:2016tdf, Armas:2018atq, Glorioso:2018kcp}. The dual formulation makes all the global symmetries of the system transparent in a model-independent manner and allow us to construct an effective field theories without the knowledge of the underlying field content, such as Goldstone fields and dynamical gauge fields in the examples above.

In the context of this work, a system with spontaneously broken $U(1)^{(q)} \times_{\kappa} U(1)^{(2q+1)}$ global symmetry admits a dual description in terms of an anomalous extended higher-group symmetry involving four higher-form symmetries. In detail, we can identify the dual higher-form currents
\begin{align}
\label{eq:dual-currents}
    \tilde J_{p+1} 
    &= \star F_{q+1}~~, \nn\\
    \tilde\cJ_{p-q} 
    &= \star\!\lb \cF_{2q+2} - \frac{\tilde\kappa}{c_\phi} F_{q+1}\wedge F_{q+1}\rb~~,
\end{align}
where $p=d-2-q$ and $\tilde\kappa = \half \kappa c_\Phi/c_\phi$. In terms of these, the Bianchi identities associated with the Goldstones $\phi_q$ and $\Phi_{2q+1}$ are recast into higher-form conservation laws
\begin{align}
    \label{eq:dual-conservation}
    \df{\star \tilde J}_{p+1} 
    &= (-)^{p} c_\phi G_{q+2}~~, \nn\\
    \df{\star \tilde\cJ_{p-q}}
    &= - c_\Phi \cG_{2q+3} 
    - (-)^{p} 2\tilde\kappa\, G_{q+2}\wedge {\star\tilde J_{p+1}}~~,
\end{align}
that, together with the original higher-group conservation laws in \cref{conservationlaw}, give rise to an extended higher-group structure 
\begin{equation}
    \bigg(
    \Big(U(1)^{(q)}\times_\kappa
    U(1)^{(2q+1)}\Big)
    \times\tilde U(1)^{(p-q-1)}\bigg)
    \times_{\tilde\kappa} \tilde U(1)^{(p)}~~.
    \label{eq:extended-higher-group}
\end{equation}
This is a $(2q+2)$-group in $d\leq 3q+3$ and $(p+1)$-group in $d\geq 3q+3$. We have used the ``tilde'' notation to distinguish the higher-form symmetries in the dual sector. In this formulation, the Goldstone charges $c_\phi$ and $c_\Phi$ appear as coefficients of mixed 't Hooft anomalies in $U(1)^{(q)}\times \tilde U(1)^{(p)}$ and $U(1)^{(2q+1)}\times \tilde U(1)^{(p-q-1)}$ sectors of the extended higher-group respectively. 

To make the extended higher-group and anomaly structure of the dual formulation manifest, let us introduce new background gauge fields $\tilde A_{p+1}$ and $\tilde B_{p-q}$ coupled to the currents $\tilde J_{p+1}$ and $\tilde\cJ_{p-q}$ via the coupling terms in the action
\begin{align}
    S
    &\sim \int 
    \lb \tilde A_{p+1} + \tilde\kappa\, A_{q+1}\wedge \tilde B_{p-q} \rb 
    \wedge {\star\tilde J}_{p+1}
    + \tilde B_{p-q} \wedge {\star\tilde\cJ_{p-q}} \nn\\
    &\sim \int 
    (-)^{p}\lb \tilde A_{p+1} + \tilde\kappa\, A_{q+1}\wedge \tilde B_{p-q} \rb
    \wedge F_{q+1} \nn\\
    &\qquad\qquad\qquad
    - \tilde B_{p-q} \wedge \lb \cF_{2q+2} 
    - \frac{\tilde\kappa}{c_\phi} F_{q+1}\wedge F_{q+1}\rb~~.
    \label{eq:coupling-terms}
\end{align}
The action is invariant under an extended higher-group symmetry
\begin{align}
    \delta A_{q+1} 
    &= \df\Lambda_{q}~~, \nn\\
    \delta B_{2q+2}
    &= \df\Lambda_{2q+1} 
    - \kappa\, \df\Lambda_{q} \wedge A_{q+1}~~, \nn\\
    \delta\tilde A_{p+1} 
    &= \df\tilde\Lambda_{p}
    + (-)^{p}\tilde\kappa\, \df\tilde\Lambda_{p-q-1} \wedge A_{q+1}
    - \tilde\kappa\, \df\Lambda_{q}\wedge \tilde B_{p-q}~~, \nn\\
    \delta\tilde B_{p-q} 
    &= \df\tilde\Lambda_{p-q-1}~~,
\end{align}
up to the residual anomaly terms 
\begin{align}
    \delta_\Lambda S 
    &= - \int (-)^{q} c_\phi G_{q+2} \wedge \tilde\Lambda_{p}
    - c_\Phi \cG_{2q+3}\wedge \tilde\Lambda_{p-q-1}
    + 2(-)^{p} c_\phi\tilde\kappa\,
    \tilde\Lambda_{p-q-1} \wedge A_{q+1}\wedge G_{q+2}~~.
\end{align}
For later use, we identify the background field strengths in the dual sector
\begin{align}
    \tilde G_{p+2}
    &= \df\tilde A_{p+1} 
    - \tilde\kappa\, \tilde B_{p-q} \wedge F_{q+2}
    - \tilde\kappa\, A_{q+1}\wedge \tilde G_{p-q+1}~~, \nn\\
    \tilde\cG_{p-q+1} 
    &= \df\tilde B_{p-q}~~.
\end{align}
The anomaly obtained above can be countered using the anomaly inflow mechanism by putting the theory on the boundary of a $(d+1)$-dimensional bulk spacetime, carrying a higher-group Chern-Simons theory with action\footnote{The associated anomaly polynomial is simply ${\cal P} = c_\phi G_{q+2}\wedge\tilde G_{p+2}
    - c_\Phi \cG_{2q+3}\wedge \tilde\cG_{p-q+1}$.}
\begin{align}
    S_{\text{bulk}}
    &= \int_{\text{bulk}}
    c_\phi G_{q+2}\wedge \tilde A_{p+1}
    + c_\Phi \lb \cG_{2q+3}
    + \frac{1}{2} \kappa\, A_{q+1}\wedge G_{q+2} \rb \wedge \tilde B_{p-q}~~.
    \label{eq:bulk-action-dual}
\end{align}
We can explicitly check that the total action $S_{\text{tot}} = S + S_{\text{bulk}}$ is invariant under the complete extended higher-group symmetry.

To identify the anomalous conservation laws associated with the extended higher-group symmetry, let us parametrise the variation of the total action $S_{\text{tot}}$ as
\begin{align}
    \delta S_{\text{tot}}
    &= \int \delta A_{q+1}\wedge {\star J_{q+1}}
    + \lb \delta B_{2q+2} + \kappa\, A_{q+1}\wedge \delta A_{q+1} \rb\wedge {\star\cJ_{2q+2}} \nn\\
    &\qquad\qquad\qquad 
    + \lb \delta\tilde A_{p+1}
    + \tilde\kappa\, A_{q+1}\wedge \delta\tilde B_{p-q}
    - \tilde\kappa\, \delta A_{q+1}\wedge \tilde B_{p-q}
    \rb
    \wedge {\star \tilde J_{p+1} }
    + \delta \tilde B_{p-q}\wedge {\star \tilde\cJ_{p-q}}
    \nn\\
    &\qquad 
    + \int_{\text{bulk}}
    \delta A_{q+1}\wedge c_\phi \tilde G_{p+2}
    - \lb \delta B_{2q+2} 
    + \kappa A_{q+1}\wedge \delta A_{q+1}  \rb \wedge c_\Phi\tilde\cG_{p-q+1}
    \nn\\
    &\qquad\qquad\qquad 
    + \lb \delta\tilde A_{p+1}
    + \tilde\kappa\, A_{q+1}\wedge \delta\tilde B_{p-q}
    - \tilde\kappa\, \delta A_{q+1}\wedge \tilde B_{p-q}
    \rb
    \wedge c_\phi G_{q+2} \nn\\
    &\qquad\qquad\qquad 
    + \delta \tilde B_{p-q}\wedge (-)^{p} c_\Phi \cG_{2q+3}~~,
    \label{eq:total-variation}
\end{align}
where the associated conserved currents have been defined to be gauge-invariant. A consequence of gauge-invariance is that the original higher-form currents $J_{q+1}$ and $\cJ_{2q+2}$ get improved compared to their original definitions with terms involving the ``tilde'' background fields, i.e.
\begin{align}
    {\star J_{q+1}}
    &\sim 
    c_\phi \tilde A_{p+1}
    + \frac{1}{2} c_\Phi\kappa\, A_{q+1}\wedge \tilde B_{p-q}
    + \Omega \tilde\kappa\, \tilde B_{p-q}
    \wedge {\star \tilde J_{p+1}}~~, \nn\\
    {\star\cJ_{2q+2}}
    &\sim c_\Phi \tilde B_{p-q}~~.
\end{align}
The last term in the first line arises from the variation of the background coupling terms in \cref{eq:coupling-terms}, while the remaining terms arise from the boundary variation of the bulk action in \cref{eq:bulk-action}. The currents follow a set of anomalous conservation laws 
\begin{align}
    \df{\star J_{q+1}} &=
    c_\phi \tilde G_{p+2}
    + 2\kappa\,G_{q+2}\wedge {\star\cJ_{2q+2}}
    + 2\tilde\kappa\,\tilde\cG_{p-q+1}\wedge {\star \tilde J_{p+1}}~~, \nn\\
    \df{\star\cJ_{2q+2}} &= c_\Phi \tilde\cG_{p-q+1}~~, \nn\\ 
    \df{\star \tilde J_{p+1}}
    &= (-)^{p} c_\phi G_{q+2}~~, \nn\\
    \df{\star \tilde\cJ_{p-q}}
    &= - c_\Phi\cG_{2q+3} 
    - (-)^{p}2\tilde\kappa\, G_{q+2}\wedge {\star\tilde J_{p+1}}~~,
\end{align}
manifesting the extended higher-group structure \eqref{eq:extended-higher-group}, with mixed anomalies $c_\phi$ in $U(1)^{(q)}\times\tilde U(1)^{(p)}$ and $c_\Phi$ in $U(1)^{(2q+1)}\times\tilde U(1)^{(p-q-1)}$ sectors.

\paragraph*{Dual formulation in sub-critical dimensions:}

The situation is qualitatively different in $d=2q+2$ dimensions. As we discussed in \cref{sub-critical}, the higher-form analogue of Mermin-Wagner theorem forbids spontaneous breaking of $U(1)^{(2q+1)}$ symmetry in $d=2q+2,2q+3$ dimensions. Nonetheless, there can be quasi-long-range order in critical dimensions $d=2q+3$ mediated by a massless quasi-Goldstone field $\phi_{2q+1}$, which allows one to set up the dual description as described above with $p=q+1$. However, this description is not valid for sub-critical dimensions $d=2q+2$ or $p=q$. The invariant field strength $\cF_{2q+2}$ associated with $\Phi_{2q+1}$ is a top-form and thus the respective Bianchi identity does not give rise to a non-trivial higher-form symmetry to set up a dual description.\footnote{The respective dual current $\tilde\cJ_0$ defined via \cref{eq:dual-currents} would be a 0-form and might be interpreted as associated with a ``$(-1)$-form symmetry''; see e.g.~\cite{Cordova:2019jnf, Cordova:2019uob, Aloni:2024jpb}. However, we do not investigate this perspective in this work.}

We can set up a dual description for a theory with $U(1)^{(q)} \times_{\kappa} U(1)^{(2q+1)}$ symmetry in sub-critical dimensions in a phase where the $U(1)^{(q)}$ part of the symmetry is spontaneously broken. As we discussed in \cref{sec:M5-brane}, for $q=2$ this is the relevant framework for the effective description of a bosonic M5 brane is supergravity. In this case, we only get a $\tilde U(1)^{(q)}$ symmetry in the dual sector arising from the Bianchi identity associated with $\phi_q$. To probe the extended higher-group structure, let us modify the action in \cref{eq:action-subcritical} by coupling it to the dual gauge field $\tilde A_{q+1}$ as
\begin{align}
    \label{eq:action-subcritical-modified}
    S 
    &= \hat S[F_{q+1}] 
    + Q_{(2q+1)}\int B_{2q+2} + \frac{\kappa}{c_\phi} A_{q+1} \wedge F_{q+1} 
    + \int \tilde A_{q+1} \wedge F_{q+1}~~,
\end{align}
and couple it to a $(d+1)$-dimensional bulk anomaly inflow action featuring pure and mixed anomalies
\begin{equation}
    S_{\text{bulk}}
    = \int_{\text{bulk}}
    C\,A_{q+1} \wedge G_{q+2} 
    + \lb c_\phi - \kappa_\times Q_{(2q+1)}\rb \tilde A_{q+1} \wedge  G_{q+2}~~. \nn\\
\end{equation}
Here we have introduced two new free parameters $C$ and $\kappa_\times$ whose relevance will be clear momentarily. The action is designed so as to be invariant under a generic $(2q+2)$-group symmetry structure involving two $q$-form symmetries and a $(2q+1)$-form symmetry i.e.
\begin{align}
    \delta A_{q+1} 
    &= \df\Lambda_{q}~~, \nn\\
    \delta \tilde A_{q+1} 
    &= \df\tilde\Lambda_{q}~~, \nn\\ 
    \delta B_{2q+2}
    &= \df\Lambda_{2q+1} 
    - \lb\kappa - \frac{C}{Q_{(2q+1)}}\rb \df\Lambda_{q} \wedge A_{q+1}
    - \kappa_\times\,\df\tilde\Lambda_q\wedge A_{q+1}~~,
\end{align}
with the structure constants $\kappa$ and $\kappa_\times$. In principle, we could also introduce a structure constant analogous to $\kappa$ in the $\tilde U(1)^{(q)}$ sector by including a term proportional to $\df\tilde\Lambda_{q} \wedge \tilde A_{q+1}$ in the transformation rule of $B_{2q+2}$. However, we can always linearly redefine the $U(1)^{(q)}$ and $\tilde U(1)^{(q)}$ symmetries to remove this. Similarly, a term like $\df\Lambda_{q} \wedge \tilde A_{q+1} - \df\tilde\Lambda_q\wedge A_{q+1}$ can be removed by redefining $B_{2q+2}$ with a term proportional to $A_{q+1}\wedge\tilde A_{q+1}$. 
The variation of the total action $S_{\text{tot}} = S + S_{\text{bulk}}$ can be parameterized as
\begin{align}
    \delta S_{\text{tot}}
    &= \int \delta A_{q+1}\wedge {\star J_{q+1}}
    + \delta \tilde A_{q+1}\wedge {\star \tilde J_{q+1}} \nn\\
    &\hspace{4em}
    + \lb \delta B_{2q+2} 
    + \lb\kappa - \frac{C}{Q_{(2q+1)}}\rb A_{q+1}\wedge \delta A_{q+1}
    + \kappa_\times \tilde A_{q+1}\wedge \delta A_{q+1}
    \rb \wedge {\star\cJ_{2q+2}}
    \nn\\
    &\qquad 
    + \int_{\text{bulk}}
    \delta A_{q+1}\wedge 
    \lb 2C\,F_{q+2}+ C_\times \tilde F_{q+2} \rb 
    + \delta \tilde A_{q+1} \wedge C_\times G_{q+2} ~~,
    \label{eq:total-variation-dual}
\end{align}
defined in a way that all the currents are gauge-invariant. The associated anomalous conservation equations are given as
\begin{align}
    \df{\star J_{q+1}}
    &= c_\phi \tilde G_{q+2} + 2\kappa Q_{(2q+1)} G_{q+2} 
    ~~, \nn\\
    \df{\star\tilde J_{q+1}}
    &= c_\phi G_{q+2}~~,
    \label{eq:two-eqns}
\end{align}
together with a trivial conservation equation $\df{\star\cJ_{2q+2}}= 0$ due to $\star \cJ_{2q+2} = Q_{(2q+1)}$. 

As we highlighted in \cref{sub-critical}, the $U(1)^{(q)} \times_{\kappa} U(1)^{(2q+1)}$ higher-group structure is quite trivial in sub-critical dimensions and can effectively be viewed as an anomaly in the $U(1)^{(q)}$ symmetry. The same applies in the dual description as well. Note that the free parameters $C$ and $\kappa_\times$ do not appear in the conservation equations and can be tuned freely to arrive at different interpretations. By choosing $C = \kappa Q_{(2q+1)}$ and $\kappa_\times = 0$, the theory can be viewed as admitting a pure $U(1)^{(q)}$ anomaly $\kappa Q_{(2q+1)}$ and a mixed $U(1)^{(q)}\times\tilde U(1)^{(q)}$ anomaly $c_\phi$ without any higher-group structure. On the other hand, by tuning $C=0$ and $\kappa_\times = c_\phi/Q_{(2q+1)}$, the theory can be viewed as realizing a non-anomalous $(2q+2)$-group symmetry with structure constants $\kappa$ and $c_\phi/Q_{(2q+1)}$. This latter perspective is sensible in an embedding scenario, as is the case for an M5 brane, when the theory under consideration lives on a $(2q+2)$-dimensional worldsheet propagating in higher dimensions. More discussion along these lines is presented in \cref{sec:embedding}.

Inspecting \cref{eq:two-eqns}, we also note that we can describe a Goldstone $\phi_q$ with (anti-)self-dual field strength by simply aligning the two currents, i.e.
\begin{equation}
    \tilde J_{q+1} = \frac{c_\phi}{2\kappa Q_{(2q+1)}} J_{q+1}~~.
\end{equation}
This causes the two conservation equations in \cref{eq:two-eqns} to coincide when the dual background field strength $\tilde G_{q+2}$ is turned off. Recall our previous discussion of (anti-)self-dual limit around \cref{eq:non-linear-self-duality}. In some ways, the dual description is more natural to talk about (anti-)self-duality because it treats the Bianchi identity and the higher-form conservation equation democratically from the get-go. To this end, one might guess that there is a formulation of self-dual field theories where we impose a $(q+1)$-form diagonal-shift symmetry between the two background gauge fields $\delta A_{q+1} = \Lambda_{q+1}$, $\delta \tilde A_{q+1} = -2\kappa Q_{(2q+1)}/c_\phi\,\Lambda_{q+1}$, causing the two operators $J_{q+1}$ and $\tilde J_{q+1}$ to align. We will leave this thought with the reader to ponder over.

\vspace{1em}

As a final comment, we note that the notations for ``tilde'' quantities $\tilde n_p$, $\tilde n_{p-q-1}$ and $\tilde\mu_p$, $\tilde\mu_{p-q-1}$ in \cref{sec:phases} are motivated from the dual formulation, and are nothing but the thermodynamic densities and the associated chemical potentials respectively corresponding to the dual symmetries $\tilde U(1)^{(p)}$, $\tilde U(1)^{(p-q-1)}$. This is made evident by the Bianchi identities in \cref{eq:mu2-Bianchi,eq:np-Bianchi,eq:mu5-Bianchi,eq:n-higher-Bianchi} that take a similar form as the configuration equations \eqref{eq:config-CC} arising from the conservation laws. In fact, the equilibrium effective actions written in \cref{sec:phases} are in the grand-canonical ensemble with respect to the symmetries $U(1)^{(q)}$, $U(1)^{(2q+1)}$, but canonical ensemble with respect to the dual symmetries $\tilde U(1)^{(p)}$, $\tilde U(1)^{(p-q-1)}$. By appropriately introducing the dual background gauge fields $\tilde A_{p+1}$, $\tilde B_{p-q}$ in \cref{sec:phases}, we can also write down dual equilibrium effective actions that are in the grand-canonical ensemble with respect to all the symmetries, though one needs to be careful about the mixed anomalies. The analogous discussion for anomalous higher-form symmetries appears in~\cite{Armas:2023tyx}; the extension to higher-group symmetries is straightforward and is left for the future.

\section{Outlook}
\label{sec:outlook}

In this work we have considered effective field theories invariant under a $(2q+2)$-group symmetry, involving a $q$-form symmetry $U(1)^{(q)}$ and a $(2q+1)$-form symmetry $U(1)^{(2q+1)}$, emphasizing on their application to supergravity. We have studied different spontaneous symmetry breaking patterns that these theories can exhibit at zero and finite temperature, incorporating temporal and complete spontaneous breaking of higher-form symmetries at finite temperature. In particular, we have explored the idea that the bosonic field content on the M5 brane can be understood as arising from spontaneous breaking of global translations and 6-group symmetries. This resulted in a formulation of the theory defined on the M5 worldvolume as a 6-group invariant theory, opening the door for a complete interpretation of the M5 brane action purely in terms of global symmetries. 

The treatment of the M5 brane that was presented here is a first step towards understanding the bosonic sector of the theories on other string theory solitons as particular cases of Goldstone theories invariant under some higher-group symmetry. For instance, in order to broaden the analysis presented here as to include the D1 string, D3, D5/NS5 and D7 branes of type IIB supergravity, and all permitted bound states formed by them, the starting point is a close inspection of the type IIB modified conservation laws
\begin{align}
    \df{\star j_{2}}
    &= 0~~, \nn\\
    \df{\star J_{2}}
    &= H_{3}\wedge {\star J_{4}}
    - \tilde{F}_{5}\wedge \star j_{6}~~, \nn\\
    \df{\star J_{4}}
    &= F_{3}\wedge {\star j_{6}}
    + H_{3} \wedge {\star \mathcal{J}_{6}}~, \nn\\
    \df{\star j_{6}}
    &= 0~~, \nn\\
    \df{\star\mathcal{J}_{6}}
    &=H_{3}\wedge \star \mathcal{J}_{8}~~, \nn\\
    \df{\star \mathcal{J}_{8}}
    &=0~~.
    \label{eq:type-IIB}
\end{align}
Here, the electric current $j_{2}$ sources the NS-NS field strength $H_{3}$, while $J_{2}$, $J_{4}$ are electric currents sourcing the R-R field strengths $F_{3},\tilde{F_{5}}$. Moreover, the magnetic currents $j_{6},\mathcal{J}_{6},\mathcal{J}_{8}$ modify the Bianchi identities for the field strengths $H_{3}, F_{3},\Tilde{F}_{5}$, respectively.\footnote{For more details, see \cite{Armas:2016mes}.} When viewed as a whole, the above equations form a rather intriguing 8-group with multiple structure constants. To each non-trivial type IIB D/NS-brane theory there is an associated subset of the equations in \eqref{eq:type-IIB}, which essentially defines the continuous higher-group for each such theory. A similar analysis can be performed for type IIA supergravity. The experience with the M5 brane leaves us optimistic that this line of thinking can be extended to the fully supersymmetric case as well by appropriately generalizing the notion of a higher-group to include fermionic symmetries. We leave the exploration of this important research line for the future.

The self-duality of the 2-form Goldstone $\phi_2$, as mentioned in section \ref{sec:M5-brane}, raises obstructions to the formulation of the M5 brane action. Proposals to deal with this problem require abandoning the manifest worldvolume covariance \cite{Perry:1996mk} or introducing additional auxiliary fields \cite{Pasti:1997gx}. In this paper, by invoking the 6-group symmetry, and then following an approach in the spirit of \cite{Witten:1996hc}, we formulated an effective action for the M5 brane that is compatible with a nonlinear self-duality constraint. However, the approach presented in this paper does not provide a symmetry-based perspective to construct actions consistent with self-duality directly.
In particular, the self-duality constraint implies that the 2-form conservation equation and the Bianchi identity are equivalent, leading to a single 2-form conservation law without an associated Bianchi identity. Correspondingly, the electric and magnetic components of the 3-form field strength $F_3$ become interdependent, giving rise to a single 2-form dynamical density. At finite temperature this setup precisely gives rise to the T-U phase, allowing us to write down an equilibrium effective action that is by construction consistent with the self-duality requirement. But it is currently unclear how to extend this understanding to the zero temperature effective action. Another possible avenue of exploration in this direction is based on the dual formulation discussed in \cref{sec:dual}, which treats the 2-form conservation equation and the associated Bianchi identity, which can also be recast as a 2-form conservation, on the same footing. In this context, one could imagine imposing a diagonal 3-form global shift symmetry between the associated background gauge fields $A_3$ and $\tilde A_3$, forcing the associated operators $J_3$ and $\tilde J_3 = \star F_3$ to align. It would be interesting to explore these ideas in more detail.

From the point of view of the low-energy description, the work carried out in this paper is an important step towards a complete classification of theories belonging to the hydrodynamic sector of supergravities. In this work, we have restricted ourselves to various higher-group phases in thermal equilibrium. An immediate step forward will be to formulate theories of higher-group hydrodynamics applicable to these phases, allowing us to describe non-equilibrium fluctuations. One may go beyond the ideal order analysis presented here and consider higher-derivative corrections, including dissipative corrections, to the higher-group constitutive relations. Returning to supergravity, such developments can be useful to better understand the dynamics and stability of thermal states and their dual (extremal) black brane configurations that break a certain number of supersymmetries \cite{Armas:2018rsy, Armas:2019asf, Armas:2022bkh}. One may also hope to be able to construct effective actions for thermal brane configurations in supergravity using the framework of Schwinger-Keldysh effective field theory~\cite{Grozdanov:2013dba, Harder:2015nxa, Crossley:2015evo, Haehl:2018lcu, Jensen:2017kzi, Jain:2020vgc, Armas:2020mpr, Jain:2020zhu, Jain:2020hcu, Jain:2023obu} that inherently implement the self-duality requirement similar to the equilibrium thermal effective actions presented in this work. We plan to return to these ideas in a future publication.

\acknowledgements

We would like to thank Niko Jokela, Alexandros Kanargias, Niels Obers, Napat Poovuttikul, Stathis Vitouladitis, and Ziqi Yan for various helpful discussions. The authors are partly supported by the Dutch Institute for Emergent Phenomena (DIEP) cluster at the University of Amsterdam and JA via the DIEP programme
Foundations and Applications of Emergence (FAEME). The work of AJ was partly funded by the European Union’s Horizon 2020 research and innovation programme under the Marie Skłodowska-Curie grant agreement NonEqbSK No.\,101027527. Part of this project was carried out during the ``Hydrodynamics at All Scales'' workshop at the Nordic Institute for Theoretical Physics (NORDITA), Stockholm.

\bibresources{mySpiresCollaboration_JAAJ,references}
\makereferences 

\appendix 

\end{document}